\documentclass[conference,a4paper]{IEEEtran}
 \usepackage[svgnames]{xcolor}

\usepackage{fancyhdr}
\newcommand{\co}{$\textrm{CO}_2$}
\usepackage{cite}

\usepackage{etoolbox}
\usepackage{tikz}
\usepackage{enumitem}

\newrobustcmd*{\mysquare}[1]{\tikz{\filldraw[draw=#1,fill=#1] (0,0)j
rectangle (0.2cm,0.2cm);}}

\newrobustcmd*{\mycircle}[1]{\tikz{\filldraw[draw=#1,fill=#1] (0,0) circle [radius=0.1cm];}}

\newrobustcmd*{\mytriangle}[1]{\tikz{\filldraw[draw=#1,fill=#1] (0,0) --
(0.2cm,0) -- (0.1cm,0.2cm);}}

\usepackage[T1]{fontenc}
\usepackage[utf8]{inputenc}

\usepackage{cite}
\usepackage[cmex10]{amsmath}
\usepackage{amssymb,amsfonts}
\usepackage{array}
\usepackage[caption=false,font=footnotesize]{subfig}
\usepackage{dblfloatfix}

\usepackage{algorithmic}
\usepackage{graphicx}
\graphicspath{{graphics/}{moregraphics/}}
\DeclareGraphicsExtensions{.pdf,.PDF,.jpeg,.JPEG,.jpg,.JPEG,.png,.PNG}

\usepackage{booktabs}

\usepackage{siunitx}

\usepackage{multirow}

\usepackage{textcomp}
\usepackage{xcolor}
\usepackage[bookmarks=true,breaklinks=true,colorlinks,linkcolor=black,citecolor=blue,urlcolor=black]{hyperref}

\def\BibTeX{{\rm B\kern-.05em{\sc i\kern-.025em b}\kern-.08em
    T\kern-.1667em\lower.7ex\hbox{E}\kern-.125emX}}

\begin{document}

\title{Chasing Carbon: The Elusive Environmental Footprint of Computing}

\author{}
\author{Udit Gupta$^{1,2}$, Young Geun Kim$^3$, Sylvia Lee$^2$, Jordan Tse$^2$,\\ Hsien-Hsin S. Lee$^2$, Gu-Yeon Wei$^1$, David Brooks$^{1}$, Carole-Jean Wu$^2$ \\ \\ $^1$Harvard University, $^2$Facebook Inc., $^3$Arizona State University \\ \\
ugupta@g.harvard.edu \,\,\, carolejeanwu@fb.com}

\maketitle

\begin{abstract}
Given recent algorithm, software, and hardware innovation, computing has enabled a plethora of new applications. As computing becomes increasingly ubiquitous, however, so does its environmental impact.
This paper brings the issue to the attention of computer-systems researchers. 
Our analysis, built on industry-reported characterization, quantifies the environmental effects of computing in terms of carbon emissions. Broadly, carbon emissions have two sources: operational energy consumption, and hardware manufacturing and infrastructure. Although carbon emissions from the former are decreasing thanks to algorithmic, software, and hardware innovations that boost performance and power efficiency, the overall carbon footprint of computer systems continues to grow.
This work quantifies the carbon output of computer systems to show that most emissions related to modern mobile and data-center equipment come from hardware manufacturing and infrastructure.
We therefore outline future directions for minimizing the environmental impact of computing systems.

\end{abstract}


\section{Introduction}
The world has seen a dramatic advancement of information and communication technology (ICT).
The rise in ICT has resulting in a proliferation of consumer devices (e.g., PCs, mobile phones, TVs, and home entertainment systems), networking technologies (e.g., wired networks and 3G/4G LTE), and data centers.
Although ICT has enabled applications including cryptocurrencies, artificial intelligence (AI), e-commerce, online entertainment, social networking, and cloud storage, it has incurred tremendous environmental impacts. 

\begin{figure}[t!]
  \centering
  \includegraphics[width=\columnwidth]{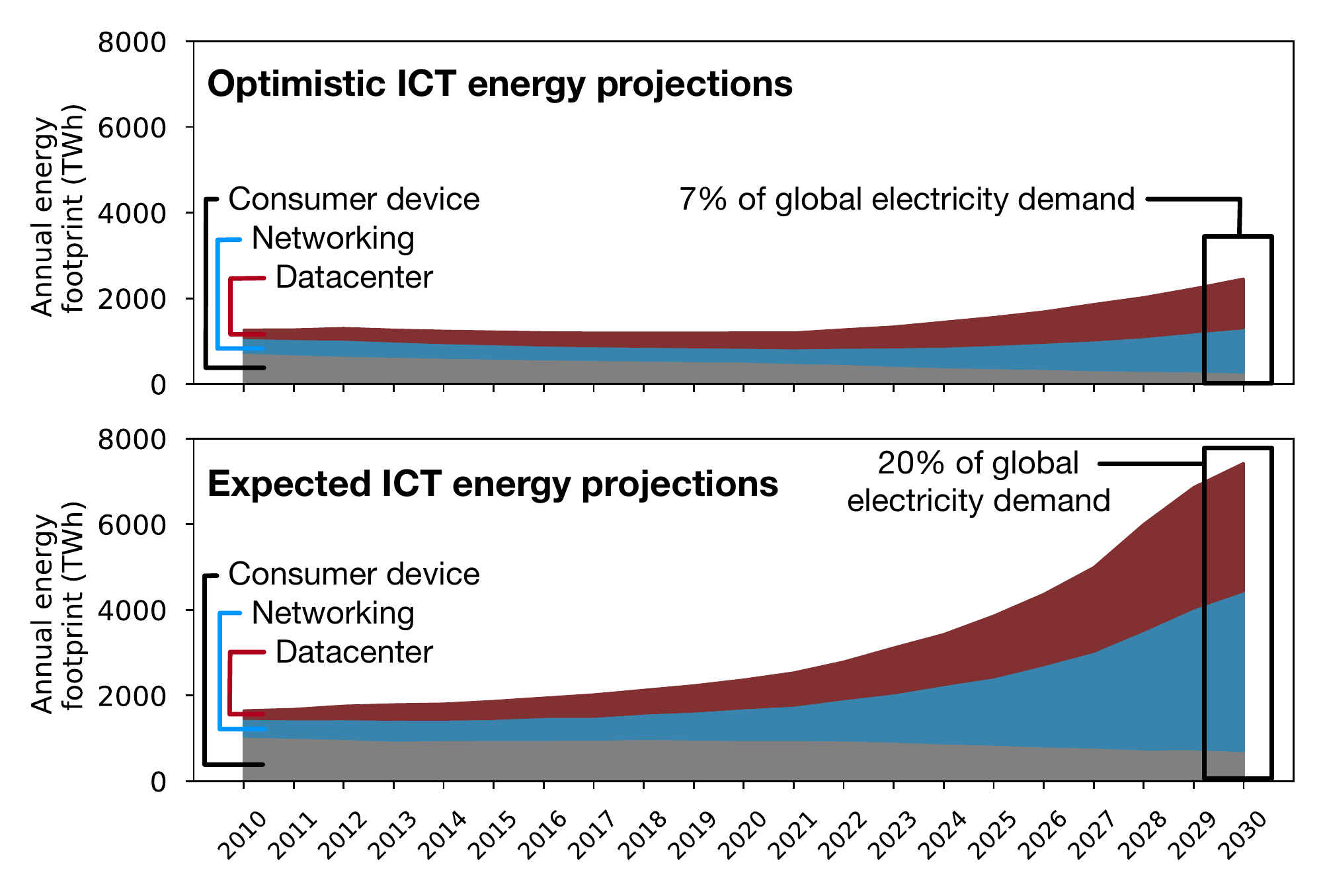}
  \vspace{-2em}
  \caption{Projected growth of global energy consumption by information and computing technology (ICT). 
  On the basis of optimistic (top) and expected (bottom) estimates, ICT will by 2030 account for 7\% and 20\% of global demand, respectively~\cite{andrae2015global}. }
  \label{fig:motivation}
  \vspace{-1.5em}
\end{figure}

Figure~\ref{fig:motivation} shows that ICT is consuming more and more electricity worldwide.
The data shows both optimistic (top) and expected (bottom) trends across mobile, networking, and data-center energy consumption~\cite{andrae2015global,jones2018stop}.
On the basis of even optimistic estimates in 2015, ICT accounted for up to 5\% of global energy demand~\cite{andrae2015global,jones2018stop}.
In fact, data centers alone accounted for 1\% of this demand, eclipsing the total energy consumption of many nations.
By 2030, ICT is projected to account for 7\% of global energy demand.
Anticipating the ubiquity of computing, researchers must rethink how to design and build sustainable computer systems.

Given the growing energy demand of computing technology, software and hardware researchers have invested heavily in maximizing the energy efficiency of systems and workloads.
For instance, between the late twentieth and early twenty-first centuries, Moore's Law has enabled fabrication of systems that have billions of transistors and 1,000$\times$ higher energy efficiency~\cite{lewisExponential}.
For salient applications, such as AI~\cite{eie,eyeriss,tpu,minerva,cambricon-x,zhang2016cambricon}, molecular dynamics~\cite{shaw2008anton}, video encoding~\cite{hameed2010understanding}, and cryptography~\cite{caulfield2016cloud}, systems now comprise specialized hardware accelerators that provide orders-of-magnitude higher performance and energy efficiency.
Moreover, data centers have become more efficient by consolidating equipment into large, warehouse-scale systems and by reducing cooling and facility overhead to improve power usage effectiveness (PUE)~\cite{barroso2009datacenter}.

\begin{figure}[t!]
  \centering
  \includegraphics[width=\columnwidth]{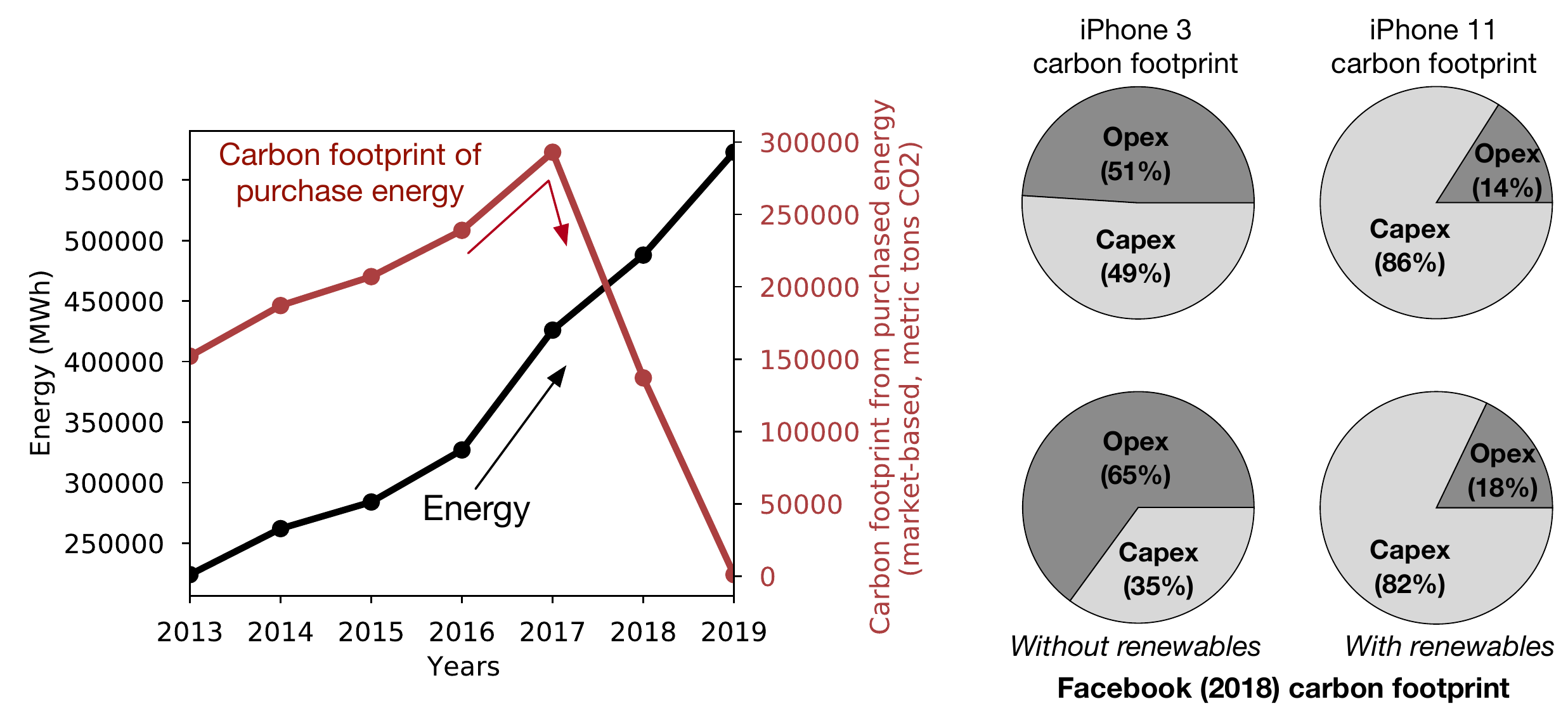}
  \vspace{-2em}
  \caption{ Carbon footprint depends on more than just energy consumption (left). Although the energy consumption of Facebook's Prineville data center increased between 2013 and 2019, its operational carbon output decreased because of renewable-energy purchases.
  The carbon-emission breakdown has shifted from primarily opex-related activities to overwhelmingly capex-related activities (right). 
  The top two pie charts show the breakdown for the iPhone 3 (2008) versus the iPhone 11 (2019); the bottom two show the breakdown for Facebook's data centers with and without renewable energy.}
  \label{fig:motivation2}
  \vspace{-1em}

\end{figure}

The aforementioned energy-efficiency improvement reduces the operational energy consumption of computing equipment, in turn mitigating environmental effects~\cite{henderson2020climate, strubell2019energy}.
In addition, using renewable energy further reduces operational carbon emissions.
Figure~\ref{fig:motivation2} (left) shows the energy consumption (black) and carbon footprint from purchased energy (red) for Facebook's Prineville data center.
Between 2013 and 2019, as the facility expanded, the energy consumption monotonically increased. On the other hand, the carbon emissions started decreasing in 2017~\cite{fbSustainabilityReport}.
By 2019, the data center's operational carbon output reached nearly zero~\cite{fbSustainabilityReport}, a direct result of powering it with green, renewable energy such as wind and solar. 
The distinction between energy consumption and carbon footprint highglights the need for computer systems to directly minimize directly optimize for environmental impact.

Given the advancements in system energy efficiency and the increasing use of renewable energy, most carbon emissions now come from infrastructure and the hardware.
Similar to dividing data-center infrastructure-efficiency optimization into opex (recurring operations) and capex (one-time infrastructure and hardware), we must divide carbon emissions into opex- and capex-related activities.
For the purposes of this work, we define opex-related emissions as emissions from hardware use and operational energy consumption; we define capex-related emissions as emissions from facility-infrastructure construction and chip manufacturing (e.g., raw-material procurement, fabrication, packaging, and assembly. 
Figure~\ref{fig:motivation2} (top, right) shows that between 2009 (iPhone 3GS) and 2019 (iPhone 11), most carbon emissions attributable to mobile devices shifted from opex related to capex related~\cite{iphone11Env, iphone3GS}.
Similarly, Figure~\ref{fig:motivation2} (bottom, right) shows that in 2018, after having converted its data centers to renewable energy, most of Facebook's remaining emissions are capex related~\cite{fbSustainabilityReport}. 

If left unchecked, we anticipate the gap between opex- and capex-related carbon output will widen in coming years.
As energy efficiency rises along with the use of renewable energy, opex-related emissions will become a less significant part of computing's environmental impact.
Increasing application demand will exacerbate capex-related emissions, however.
In less than two years, Facebook hardware devoted to AI training and inference has grown by 4$\times$ and 3.5$\times$, respectively~\cite{park2018deep, naumov2020deep}.
Likewise, to support emerging applications (e.g., AI and AR/VR) on mobile devices, smartphones today have more transistors and specialized circuits than their predecessors; limited by dark silicon~\cite{esmaeilzadeh2012dark}, the additional hardware exacerbates capex-related carbon footprints.
Addressing both opex- and capex-related emissions requires fundamentally rethinking designs across the entire computing stack.






This paper takes a data-driven approach to studying the carbon breakdown of hardware life cycle---including manufacturing, transport, use, and recycling---for consumer devices and data-center systems.
It lays the foundation for characterizing and creating more-sustainable designs.
First, we present the state of industry practice using the Greenhouse Gas (GHG) Protocol to quantify the environmental impact of industry partners and to study the carbon footprint of mobile and data-center hardware (Section~\ref{sec:quantify_carbon}).
On the basis of publicly available sustainability reports from AMD, Apple, Facebook, Google, Huawei, Intel, Microsoft, and TSMC, we show that the hardware-manufacturing process, rather than system operation, is the primary source of carbon emissions (Section~\ref{sec:mobile} and ~\ref{sec:datacenters}).
Despite the growing use of renewable energy to power semiconductor manufacturing, hardware manufacturing and capex-related activities will continue to dominate the carbon output~(Section~\ref{sec:manufacturing}). 
Finally, we outline future research and design directions across the computing stack that should enable the industry to realize environmentally sustainable systems and to reduce the carbon footprint from technology (Section~\ref{sec:future_direction}).
\begin{table*}[t!]
\begin{center}
\small
\begin{tabular}{|c|c|c|c|}
\hline
\textbf{Technology company} & Scope 1 & Scope 2 & Scope 3 \\ \hline \hline 
Chip manufacturer & Burning PFCs, chemicals, gases & Energy for fabrication & Raw materials, hardware use \\ \hline
Mobile-device vendor & Natural gas, diesel & Energy for offices & Chip manufacturing, hardware use \\ \hline
Data-center operator & Natural gas, diesel & Energy for data centers &  Server-hardware manufacturing, construction \\ \hline

\end{tabular}
\end{center}
\caption{Important features of Scope 1, Scope 2, and Scope 3 emissions, following the Greenhouse Gas (GHG) Protocol, for semiconductor manufacturers, mobile-device vendors, and data-center operators. }
\vspace{-2em}
\label{tab:scopes}
\end{table*}

The important contributions of this work are: 

\begin{enumerate}[leftmargin=*]

\item We show that given the considerable efforts over the past two decades to increase energy efficiency, the dominant factor behind the overall carbon output of computing has shifted from operational activities to hardware manufacturing and system infrastructure. 
Over the past decade, the fraction of life-cycle carbon emissions due to hardware manufacturing increased from 49\% for the iPhone 3GS to 86\% for the iPhone 11.



\item Our smartphone-based measurement shows that efficiently amortizing the manufacturing carbon footprint of a Google Pixel 3 smartphone requires continuously running MobileNet image-classification inference for three years---beyond the typical smartphone lifetime. 
This result highlights the environmental impact of system manufacturing and motivates leaner systems as well as longer system lifetimes where possible.

\item We show that because an increasing fraction of warehouse-scale data centers employ renewable energy (e.g., solar and wind), data-center carbon output is also shifting from operation to hardware design/manufacturing and infrastructure construction. 
In 2019, for instance, capex- and supply-chain-related activities accounted for 23$\times$ more carbon emissions than opex-related activities at Facebook.

\item We chart future paths for software and hardware researchers to characterize and minimize computing technology's environmental impact.
Sustainable computing will require interdisciplinary efforts across the computing stack.





\end{enumerate}

\section{Quantifying environmental impact}~\label{sec:quantify_carbon}
The environmental impact of ICT is complex and multifaceted.
For instance, technology companies consider many environmental matters including consumption of energy, water, and materials such as aluminum, cobalt, copper, glass, gold, tin, lithium, zinc, and plastic. 
In this paper we focus on a single important environmental issue: carbon emissions, which represents total greenhouse-gas (GHG) emissions.

This section details the state-of-the-art industrial practices for quantifying carbon emissions.
Our discussion presents carbon-footprint-accounting methods that serve broadly across technology companies, including AMD, Apple, Facebook, Google, Huawei, Intel, Microsoft, and TSMC~\cite{globalFoundriesSustainabilityReport, intelSustainabilityReport, appleSustainabilityReport, fbSustainabilityReport, amdSustainabilityReport, googSustainabilityReport}.
First we review methods for analyzing organization-level emissions.
Next, we analyze how to use the results of such analyses across the technology supply chain to develop models for individual computer systems, including data-center and mobile platforms. The remainder of the paper builds on these methods to quantify the carbon output of computer systems.

\subsection{Industry-level carbon-emission analysis}
\begin{figure}[t!]
  \centering
  \includegraphics[width=0.95\columnwidth]{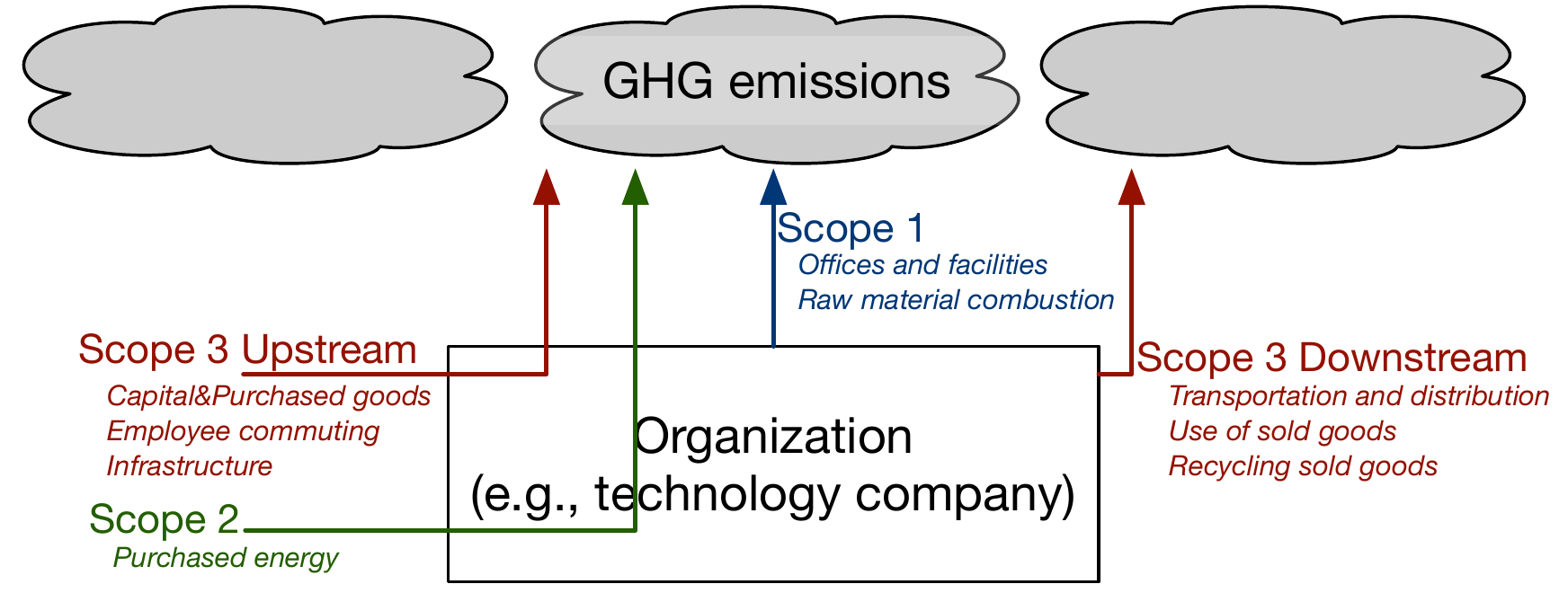}
  \vspace{-1em}
  \caption{Many organizations follow the Greenhouse Gas (GHG) Protocol to quantify their environmental impact. This protocol categorizes emissions into Scope 1 (direct), Scope 2 (indirect), and Scope 3 (upstream and downstream supply chain). }
  \label{fig:ghg_protocol}
  \vspace{-1em}
\end{figure}

A common method for quantifying organization-level carbon output is the Greenhouse Gas (GHG) Protocol~\cite{ghgprotocol}, an accounting standard by which many companies report their carbon emissions.
For example, AMD, Apple, Facebook, Google, Huawei, Intel, and Microsoft publish annual sustainability reports using the GHG Protocol~\cite{ghgprotocol}. 
Our analysis builds on such publicly available reports.
As Figure~\ref{fig:ghg_protocol} shows, the GHG Protocol categorizes emissions into three scopes: Scope 1 (direct emissions), Scope 2 (indirect emissions from purchased energy), and Scope 3 (upstream and downstream supply-chain emissions). 
We define each one in the context of technology companies, as follows. 

\textbf{Scope 1} emissions come from fuel combustion (e.g., diesel, natural gas, and gasoline), refrigerants in offices and data centers, transportation, and the use of chemicals and gases in semiconductor manufacturing.
Although Scope 1 accounts for a small fraction of emissions for mobile-device vendors and data-center operators, it comprises a large fraction for chip manufacturers.
Overall, it accounts for over half the operational carbon output from Global Foundries, Intel, and TSMC~\cite{intelSustainabilityReport,tsmcSustainabilityReport,globalFoundriesSustainabilityReport}.
Much of these emissions come from burning perfluorocarbons (PFCs), chemicals, and gases. 
TSMC reports that nearly 30\% of emissions from manufacturing 12-inch wafers are due to PFCs, chemicals, and gases~\cite{tsmcSustainabilityReport}.
In this paper we show that chip manufacturing, as opposed to hardware use and energy consumption, accounts for most of the carbon output attributable to hardware systems.

\textbf{Scope 2} emissions come from purchased energy and heat powering semiconductor fabs, offices, and data-center operation.
They depend on two parameters: the energy that operations consume and the GHG output from generating the consumed energy (in terms of carbon intensity---i.e., grams of \co~emitted per kilowatt-hour of energy).
Scope 2 emissions are especially important in semiconductor fabs and data centers.

Semiconductor companies need copious energy to manufacture chips.
Energy consumption, for instance, produces over 63\% of the emissions from manufacturing 12-inch wafers at TSMC~\cite{tsmcSustainabilityReport}.
And energy demand is expected to rise, with next-generation manufacturing in a 3nm fab predicted to consume up to 7.7 billion kilowatt-hours annually~\cite{tsmcFab3nm, tsmcSustainabilityReport}.
TSMC's renewable-energy target will account for 20\% of its fabs' annual electricity consumption, reducing its average carbon intensity. 
Despite these improvements, this work shows hardware manufacturing will constitute a large portion of computing's carbon footprint.

Scope 2 emissions are also especially important for data centers. 
The operational footprint of a data center has two parameters: the overall energy consumption from the many servers and the carbon intensity of that energy.
Note that the carbon intensity varies with energy source and grid efficiency.
Compared with ``brown'' energy from coal or gas, ``green'' energy from solar, wind, nuclear, or hydropower produces up to 30$\times$ fewer GHG emissions~\cite{weissbach2013energy, nrealpvepbt, bonou2016life}.
Scope 2 emissions for a data center therefore depend on the geographic location and energy grid.
In fact, warehouse-scale data centers are purchasing renewable energy (e.g., solar and wind) to reduce GHG emissions.

\textbf{Scope 3} emissions come from all other activities, including the full upstream and downstream supply chain.
They often comprise employee business travel, commuting, logistics, and capital goods.
For technology companies, however, a crucial and challenging aspect of carbon footprint analysis is accounting for the emissions from hardware bought and sold.
Data centers, for instance, may contain thousands of server-class CPUs whose production releases GHGs from semiconductor fabs.
Constructing these facilities also produces GHG emissions.
Similarly, mobile-device vendors must consider both the GHGs from manufacturing hardware (upstream supply chain) and the use of that hardware (downstream supply chain).
Accurate accounting of Scope 3 emissions requires in-depth analysis of GHGs resulting from construction, hardware manufacturing, and the devices' frequency of use as well as mixture of workloads.
These characteristics must consider the lifetime of systems;
for example, data centers typically maintain server-class CPUs for three to four years~\cite{barroso2009datacenter}.

Table~\ref{tab:scopes} summarizes the salient emissions from each category for chip manufacturers, mobile vendors, and data-center operators.

\begin{figure}[t!]
  \centering
  \includegraphics[width=\columnwidth]{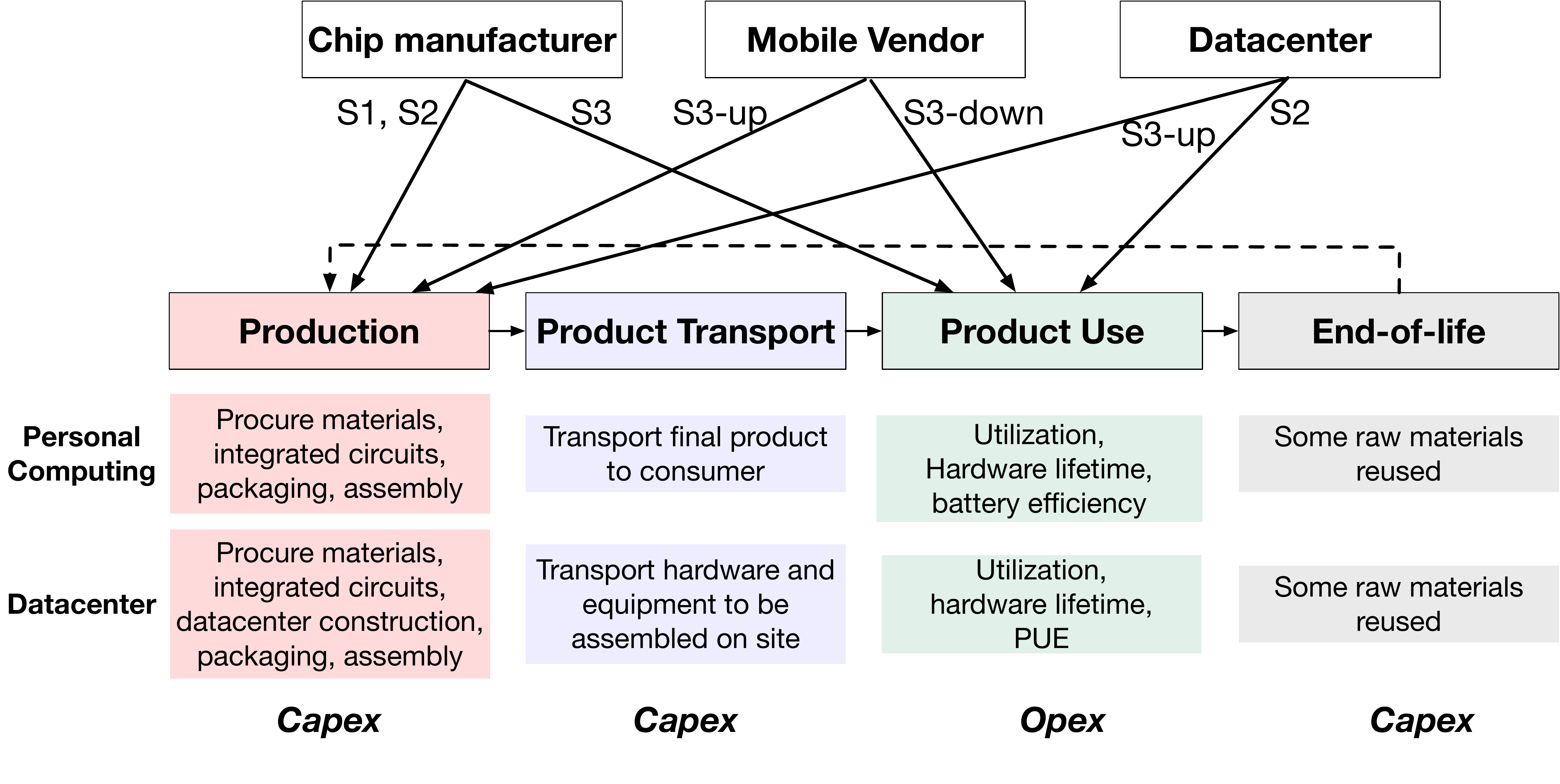}
  \vspace{-2.5em}
  \caption{ The hardware life cycle includes production, transport, use, and end-of-life processing.
  Opex-related (operational) carbon emissions are based on use; capex-related emissions results are from aggregating production/manufacturing, transport, and end-of-life processing.}
  \label{fig:life_cycle}
  \vspace{-1em}
\end{figure}

\subsection{System-level carbon-output analysis}
In addition to industry- and organization-level analysis using the GHG Protocol, carbon output can be computed for individual hardware systems and components.
Knowing the carbon footprint of individual hardware systems (e.g., server-class CPUs, mobile phones, wearable devices, and desktop PCs) not only enables consumers to understand personal carbon impact but also enables designers to characterize and optimize their systems for environmental sustainability.
Typically, evaluating the carbon footprint of an individual hardware system involves life-cycle analyses (LCAs)~\cite{ayres1995life}, including production/manufacturing, transport, use, and end-of-life processing, as Figure~\ref{fig:life_cycle} shows.

Mobile and data-center devices integrate components and IP from various organizations.
The design, testing, and manufacture of individual components (e.g., CPUs, SoCs, DRAM, and HDD/SSD storage) spreads across tens of companies.
Furthermore, mobile devices comprise displays, batteries, sensors, and cases that contribute to their carbon footprint. 
Similarly, data centers comprise rack infrastructure, networking, and cooling systems; their construction is yet another factor.
Quantifying individual systems requires quantifying GHG emissions across chip manufacturers, mobile vendors, and data-center operators.
Figure~\ref{fig:life_cycle} ties the Scope 1 (S1), Scope 2 (S2), Scope 3 upstream (S3-up), and Scope 3 downstream (S3-down) of technology companies to hardware manufacturing and operational use.
Note that although LCAs can help determine system-level emissions, they are lengthy and incur high effort to perform.

Computer systems have four LCA phases: 

\begin{itemize}[leftmargin=*]
\item \textbf{Production:} carbon emissions from procuring or extracting raw materials, manufacturing, assembly, and packaging. 

\item \textbf{Transport:} carbon emissions from moving the hardware to its point of use, including consumers and data centers. 

\item \textbf{Use:} carbon emissions from the hardware's operation, including static and dynamic power consumption, PUE overhead in the data center, and battery-efficiency overhead in mobile platforms. 

\item \textbf{End-of-life:} carbon emissions from end-of-life processing and recycling of hardware. Some materials, such as cobalt in mobile devices, are recyclable for use in future systems. 
\end{itemize}

For this paper we chose accredited and publicly reported LCAs from various industry sources, including AMD, Apple, Google, Huawei, Intel, Microsoft, and TSMC, to analyze the carbon output of computer systems.

\section{Environmental impact of personal computing}~\label{sec:mobile}
Using publicly reported carbon-emission data from industry, this section studies the environmental impact of consumer (personal) computing devices.
First, we characterize the overall carbon emissions of mobile-device developers, such as Apple, and find that hardware manufacturing dominates their environmental impact.
Next, we examine in detail various platforms (e.g., mobile phones, wearable devices, personal assistants, tablets, laptops, and desktop PCs) as well as historical trends.
Our analysis considers more than 30 products from Apple, Google, Huawei, and Microsoft.
Finally, we conduct a case study on tradeoffs between mobile performance, energy efficiency, and carbon emissions for an example AI inference workload.
The results demonstrate that software and hardware researchers should revisit mobile design to build platforms that are more efficient and environmentally sustainable.

\begin{figure}[t!]
  \centering
  \includegraphics[width=0.8\columnwidth]{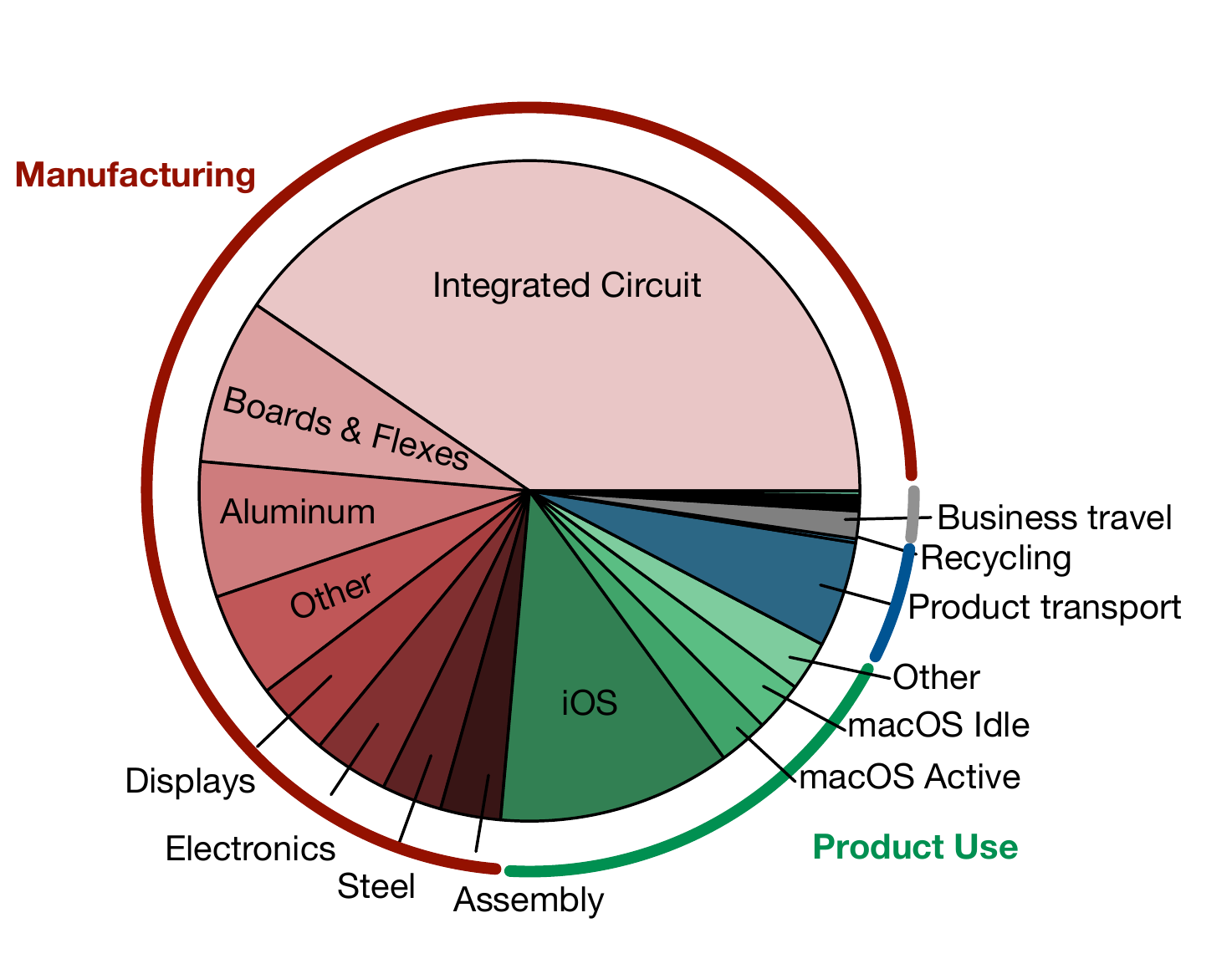}
  \vspace{-1em}
  \caption{ Apple's carbon-emission breakdown. 
  In aggregate, the hardware life cycle (i.e., manufacturing, transport, use, and recycling) comprises over 98\% of Apple's total emissions.
  Manufacturing accounts for 74\% of total emissions, and hardware use accounts for 19\%. 
  Carbon output from manufacturing integrated circuits (i.e., SoCs, DRAM, and NAND flash memory) is higher than that from hardware use.
  }
  \vspace{-1.0em}
  \label{fig:apple_breakdown}

\end{figure}

\subsection{Overall breakdown of mobile vendors}
\textbf{Takeaway 1:} \textit{Hardware manufacture and use dominate the carbon output of personal-computing companies (e.g., Apple).
More emissions come from designing and manufacturing integrated circuits (e.g., SoCs, DRAM, and storage) than from hardware use and mobile energy consumption. 
} 


Figure~\ref{fig:apple_breakdown} shows the breakdown of Apple's annual carbon footprint for 2019~\cite{appleSustainabilityReport}.
It separates the company's total emissions---25 million metric tons of \co---into manufacturing (red), product use (green), product transport (blue), corporate facilities (grey), and product recycling.
Manufacturing, which includes integrated circuits, boards and flexes, displays, electronics, steel, and assembly, accounts for over 74\% of all emissions.
By comparison, emissions from product use---energy consumption from applications running on hardware\footnote{Apple reports GHG emissions from product use on the basis of the amount of time device is in active operation and the geographically specific energy-grid efficiency.}---account for only 19\% of Apple's overall output.

Among the salient hardware-manufacturing components are integrated circuits, boards and flexes, aluminum, electronics, steel, and assembly.
Integrated circuits, comprising roughly 33\% of Apple's total carbon output, consist of CPUs, DRAMs, SoCs, and NAND flash storage~\cite{appleSustainabilityReport}.
In fact, capex-related carbon emissions from manufacturing integrated circuits alone eclipse opex-related carbon emissions from device energy consumption.
Additional capacitors, resistors, transistors, and diodes soldered to bare boards and flexes constitute ``electronics''; battery cells, plastic, and glass constitute ``other.''
The role of integrated circuits illustrates the potential impact computer-architecture and circuit researchers can have on sustainable-hardware design.

\subsection{Personal-computing life-cycle analyses}
Apple's overall carbon footprint aggregates the emissions from all of its mobile phones, tablets, wearable devices, and desktops.
Here we detail the carbon footprint of each type.
The analysis includes devices from Apple, Google, Huawei, and Microsoft.

\begin{figure}[t!]
  \centering
  \includegraphics[width=\columnwidth]{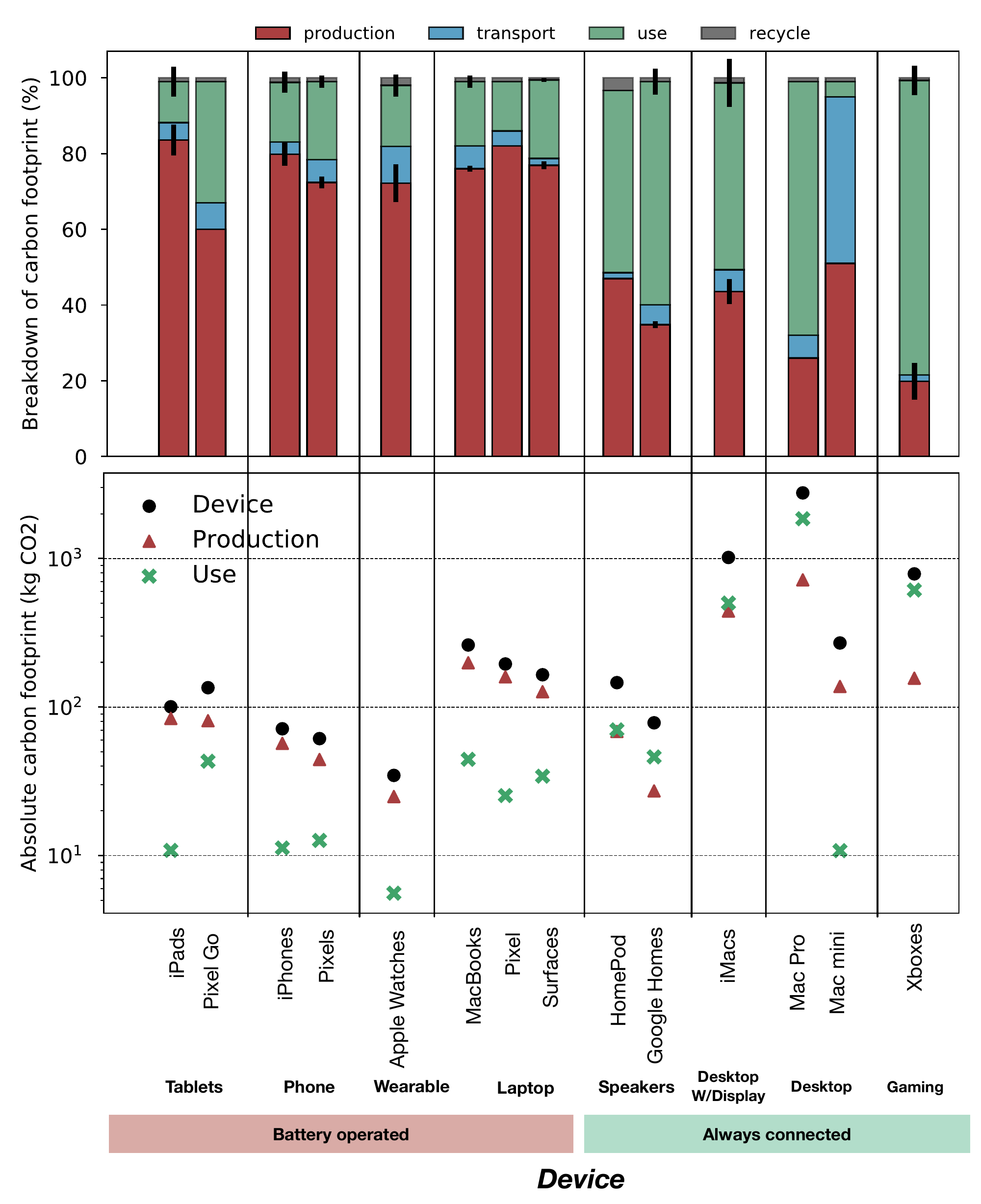}
  \vspace{-2em}
  \caption{
 Breakdown of carbon emissions for various Apple, Google, and Microsoft personal-computing platforms.
  As the top chart shows, hardware manufacturing dominates the carbon output for battery-powered devices (e.g., phones, wearables, and tables); most emissions for always connected devices (e.g., laptops, desktops, and game consoles) come from product use.
  The bottom chart shows the absolute carbon output of battery-powered and always connected devices.
  Overall, carbon footprint (total, manufacturing, and use) is variable and scales with the platform.
  }
  \label{fig:device_type_breakdown}
  \vspace{-1em}
\end{figure}

\textbf{Takeaway 2:} \textit{The breakdown of carbon footprint between manufacturing and use varies by consumer device. 
Manufacturing dominates emissions for battery-powered devices, whereas operational energy consumption dominates emissions from always-connected devices}. 

Figure~\ref{fig:device_type_breakdown} (top) shows LCAs for different battery-powered devices (e.g., tablets, phones, wearables, and laptops) and always connected devices (e.g., personal assistants, desktops, and game consoles).
The analysis aggregates LCAs from Apple, Google, and Microsoft products released after 2017~\cite{iphone11Env,iphone11ProEnv,iphone11ProMaxEnv,iphoneSEEnv,iphoneXREnv,iPadAirEnv,iPadEnv,iPadMiniEnv,iPadPro11Env,iPadPro12Env,AppleWatch3CellEnv,AppleWatch3Env,AppleWatch5Env,MacBookAirEnv,MacBookPro13Env,MacBookProEnv,MacMiniEnv,MacProEnv,Pixel2Env,Pixel2XLEnv,Pixel3aEnv,Pixel3XLEnv,Pixel3aXLEnv,Pixel3Env,GoogleHomeEnv,GoogleHomeHubEnv,GoogleHomeMiniEnv,PixelGo}. 
For devices with multiple models, such as the iPhone 11, iPhone XR, and iPhone SE, we show one standard deviation of manufacturing and operational-use breakdowns.  
For all devices, we aggregate each one's emissions across its lifetime, representing an average of three to four years for mobile phones, wearables, tablets, and desktops~\cite{appleSustainabilityReport, googSustainabilityReport}.

To reduce the carbon footprints of personal-computing devices, hardware and software designers must consider the carbon impact of both hardware manufacturing (capex) and energy consumption (opex).
For instance, Figure~\ref{fig:device_type_breakdown} (top) shows that manufacturing (capex) accounts for roughly 75\% of the emissions for battery-powered devices.
Energy consumed (opex) by these devices accounts for approximately 20\% of emissions.
By comparison, most emissions for always connected devices are from their energy consumption.
Nonetheless, even for these devices, hardware manufacturing accounts for 40\% of carbon output from personal assistants (e.g., Google Home) and 50\% from desktops.

\textbf{Takeaway 3:} \textit{In addition to the carbon breakdown, the total output for device and hardware manufacturing varies by platform.
The hardware-manufacturing footprint increases with increasing hardware capability (e.g., flops, memory bandwidth, and storage).}

Figure~\ref{fig:device_type_breakdown} (bottom) shows the absolute carbon emissions for manufacturing (\mytriangle{Crimson}), operation (\textcolor{ForestGreen}{\textbf{X}}), and the overall device total (\mycircle{black}).
Results are based on the average footprint for each device type. 

Across devices, the amount of total, manufacturing-related, and use-related emissions vary.
For instance, always connected devices typically involve more emissions than battery-powered devices.
To illustrate, the total and manufacturing footprint for an Apple MacBook laptop is typically 3$\times$ that of an iPhone.
The varying total and manufacturing levels illustrate that the capex-related output depends on the platform design and scale rather than being a static overhead. 

\textbf{Takeaway 4: } \textit{As energy efficiency improves and hardware capability increases from one device generation to another, a rising percentage of hardware life-cycle carbon emissions comes from manufacturing. }

\begin{figure}[t!]
  \centering
  \includegraphics[width=\columnwidth]{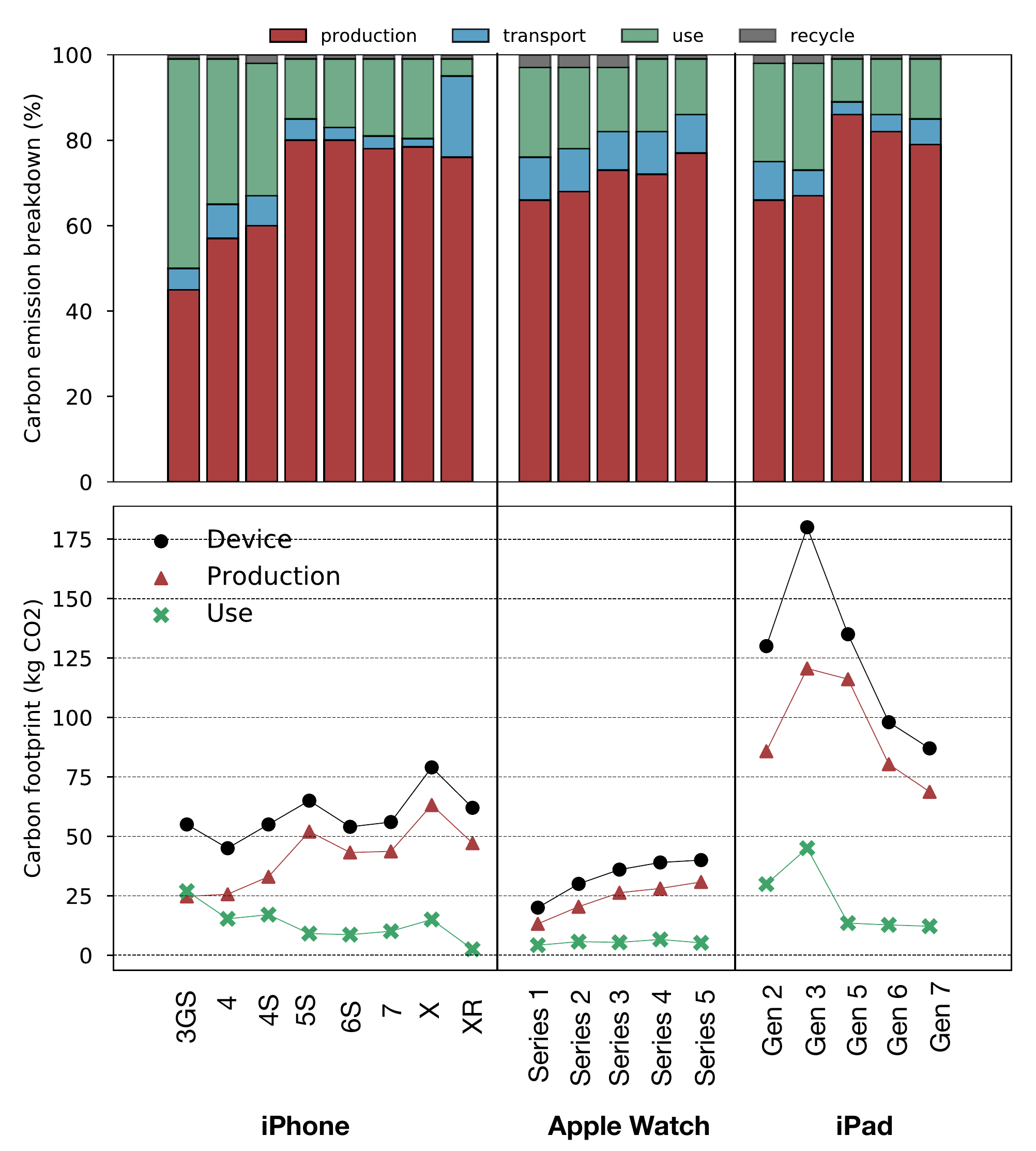}
  \vspace{-2em}
  \caption{ Carbon emissions and breakdown of emissions across generations for Apple iPhones, Watches, and iPads.
  Across all devices (top), the fraction from production and manufacturing increased from generation to generation.
  The absolute carbon output (\mycircle{black}) for iPads decreased over time, while for iPhones and Watches it increased (bottom).
  The rising carbon emissions are largely due to a growing contribution from manufacturing.
  For iPhones, as carbon from operational use (\textcolor{ForestGreen}{\textbf{X}}) decreased, the manufacturing contribution (\mytriangle{Crimson}) increased.
  }
  \vspace{-1.0em}
  \label{fig:device_generations}

\end{figure}

Figure~\ref{fig:device_generations} (top) shows the carbon breakdown over several generations of battery-powered devices: iPhones (from 2008's 3GS to 2018's XR), Apple Watches (2016's Series 1 to 2019's Series 5), and iPads (2012's Gen 2 to 2019's Gen 7).
In all three cases, the fraction of carbon emissions devoted to hardware manufacturing increased over time.
For iPhones, manufacturing accounts for 40\% of emissions in the 3GS and 75\% in the XR; for Apple Watches, it accounts for 60\% in Series 1 and 75\% in Series 5; and for iPads, 60\% in Gen2 and 75\% in Gen 7. 

Figure~\ref{fig:device_generations} (bottom) shows the absolute carbon output across generations for the same devices.
As performance and energy efficiency of both software and hardware have improved over the past few years, the opex-related carbon output from energy consumption (\textcolor{ForestGreen}{\textbf{X}}) has decreased. 
Despite the energy-efficiency increases over iPhone and Apple Watch generations, however, total carbon emissions (\mycircle{black}) grew steadily. 
The increasing ouputs owe to a rising contribution from manufacturing (\mytriangle{Crimson}) as hardware provides more flops, memory bandwidth, storage, application support, and sensors.
The opposing energy-efficiency and carbon-emission trends underscore the inequality of these two factors.
Reducing carbon output for the hardware life cycle requires design for lower manufacturing emissions or engagement with hardware suppliers.

\subsection{Performance and energy versus carbon footprint}
In addition to the overall carbon emissions from manufacturing and operational energy consumption, we also consider performance, energy, and carbon footprint tradeoffs for an example workload: mobile AI inference.

\textbf{Takeaway 5:} \textit{From 2017 to 2019, software and hardware optimizations primarily focused on maximizing performance, overlooking the growth trend of carbon footprint. 
}

\begin{figure}[t!]
  \centering
  \includegraphics[width=\columnwidth]{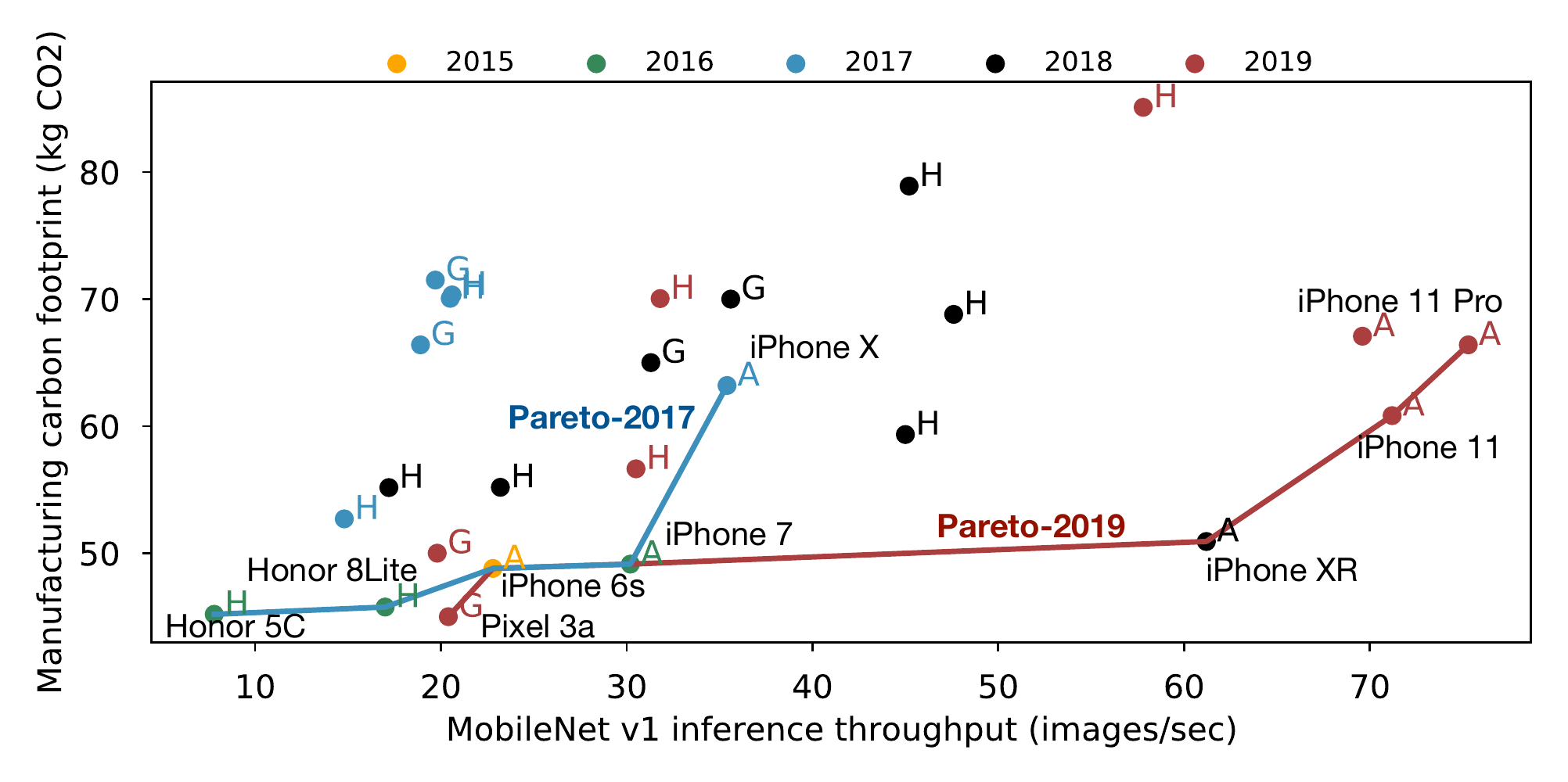}
  \vspace{-2em}
  \caption{ Performance (MobileNet v1 inference throughput) versus carbon footprint Pareto frontier by mobile-phone generation. (``A'' represents Apple, ``G'' Google, and ``H'' Huawei.)
  The Pareto frontier shifted primarily to the right between 2017 (blue line) and 2019 (red line), highlighting the focus on increasing system performance as opposed to decreasing carbon emissions.}
  \vspace{-1.0em}
  \label{fig:phone_perf_cf_pareto}

\end{figure}

Figure~\ref{fig:phone_perf_cf_pareto} illustrates the tradeoff between performance, measured as MobileNet v1 throughput (i.e., inference images per second)~\cite{geekbench,howard2017mobilenets}, and the manufacturing carbon footprint.
The analysis categorizes devices by their release year (color) and vendor (``G'' for Google, ``H'' for Huawei, and ``A'' for Apple)~\cite{iphone11Env, iphone11ProEnv, iphone3GS, iphoneXREnv, Pixel2Env, Pixel3aEnv, Pixel3Env, Pixel3XLEnv, huaweiProducts}. 
Finally, we highlight two performance/carbon footprint Pareto frontiers for devices made in 2017 and earlier (blue) and for devices made in 2019 and earlier (red).

The Pareto frontiers illustrate a tradeoff between AI performance and carbon footprint.
Across the 2019 performance/carbon-footprint frontier, the iPhone 11 Pro achieves a MobileNet v1 throughput of 75 images per second at a manufacturing output of 66 kg of \co; in comparison, the Pixel 3a achieves an inference throughput of 20 images per second with 45 kg of \co. 
In addition to this tradeoff, the Pareto frontier between 2017 and 2019 shifts to the right, prioritizing higher performance through more-sophisticated SoCs and specialized hardware.
In fact, although the iPhone X (2017) achieved a throughput of 35 images per second at 63 kg of \co, the iPhone 11 (2019) doubled that performance at a slightly lower 60 kg of \co. 
While greater performance is important to enabling new applications and improving the user experience, moving the Pareto frontier down is also important---that is, to mitigate the environmental impact of emerging platforms and applications, it is crucial to design workloads and systems with similar performance but lower environmental impact.

\begin{figure}[t!]
  \centering
  \includegraphics[width=0.925\columnwidth]{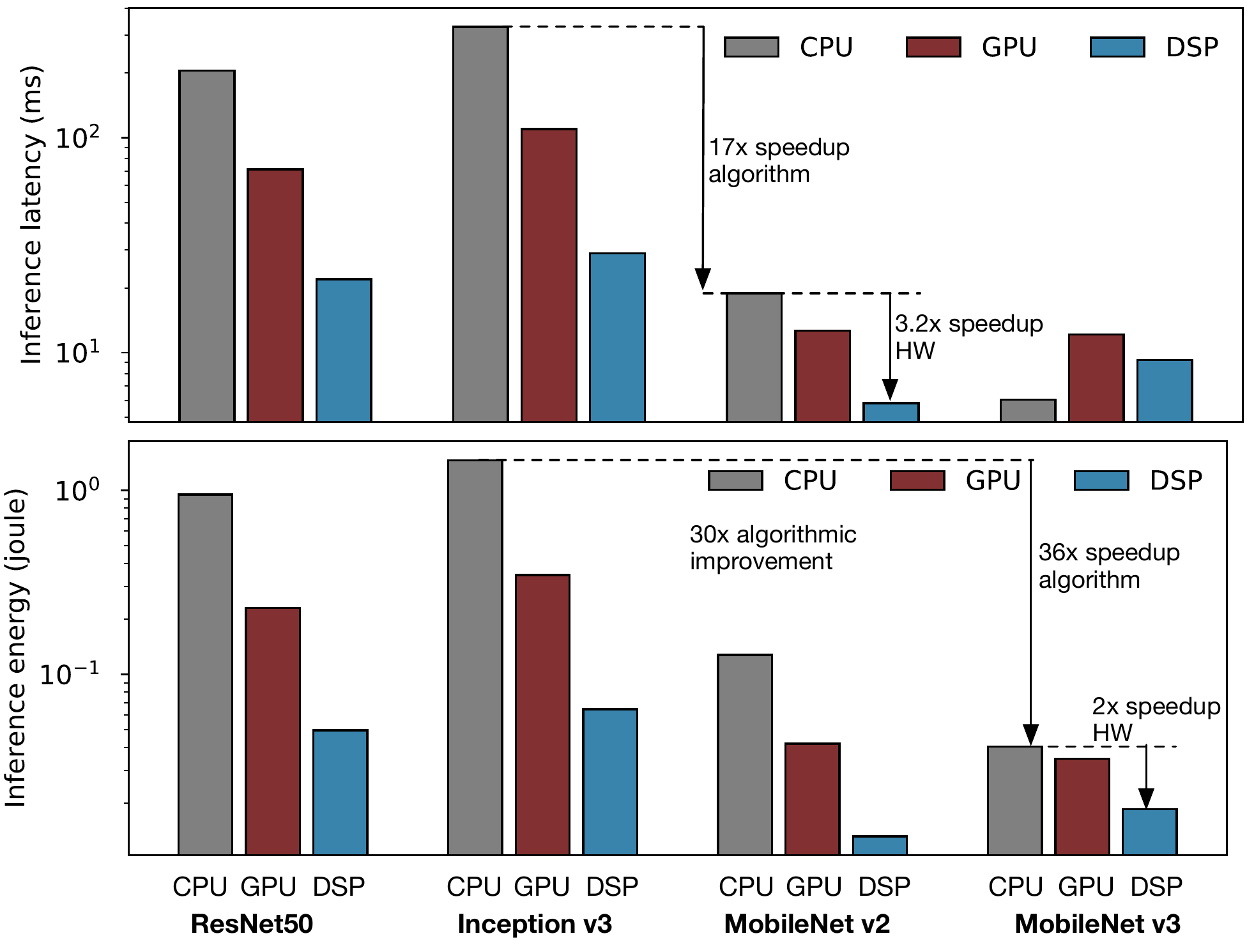}
  \vspace{-1em}
  \caption{  Evaluating the improvement of inference throughput (top) and energy efficiency (bottom) for different convolutional-neural-network and hardware generations. 
  Algorithmic and hardware advances have considerably increased the performance and operational energy consumption.}
  \vspace{-1.0em}
  \label{fig:energy_pixel3a}

\end{figure}

\textbf{Takeaway 6: } \textit{Given the energy-efficiency improvements from software and hardware innovation over the last decade, amortizing the manufacturing carbon output requires continuously operating mobile devices for three years---beyond their typical lifetime.}

Figure~\ref{fig:energy_pixel3a} illustrates the inference latency (top) and energy (bottom) of several well-known convolutional neural networks.
Results are for a unit batch size and 224$\times$224 images on a Google Pixel 3 phone with a Qualcomm Snapdragon 845 SoC~\cite{russakovsky2015imagenet}.
We measured energy consumption on a Monsoon power monitor~\cite{msoon, kim2020autoscale}.
As expected, algorithmic and hardware innovation has improved both performance and energy efficiency.
For instance, when running on a CPU, MobileNet v2 is 17$\times$ faster than Inception v3~\cite{howard2019searching}.
Moreover, it is an additional 3$\times$ faster when running on a DSP than on a CPU.
Similarly, algorithmic and hardware innovation has increased energy efficiency by 36$\times$ and 2$\times$, respectively.
The performance and energy optimizations have also affected AI carbon footprint on mobile devices.

\begin{figure}[t!]
  \centering
  \includegraphics[width=0.925\columnwidth]{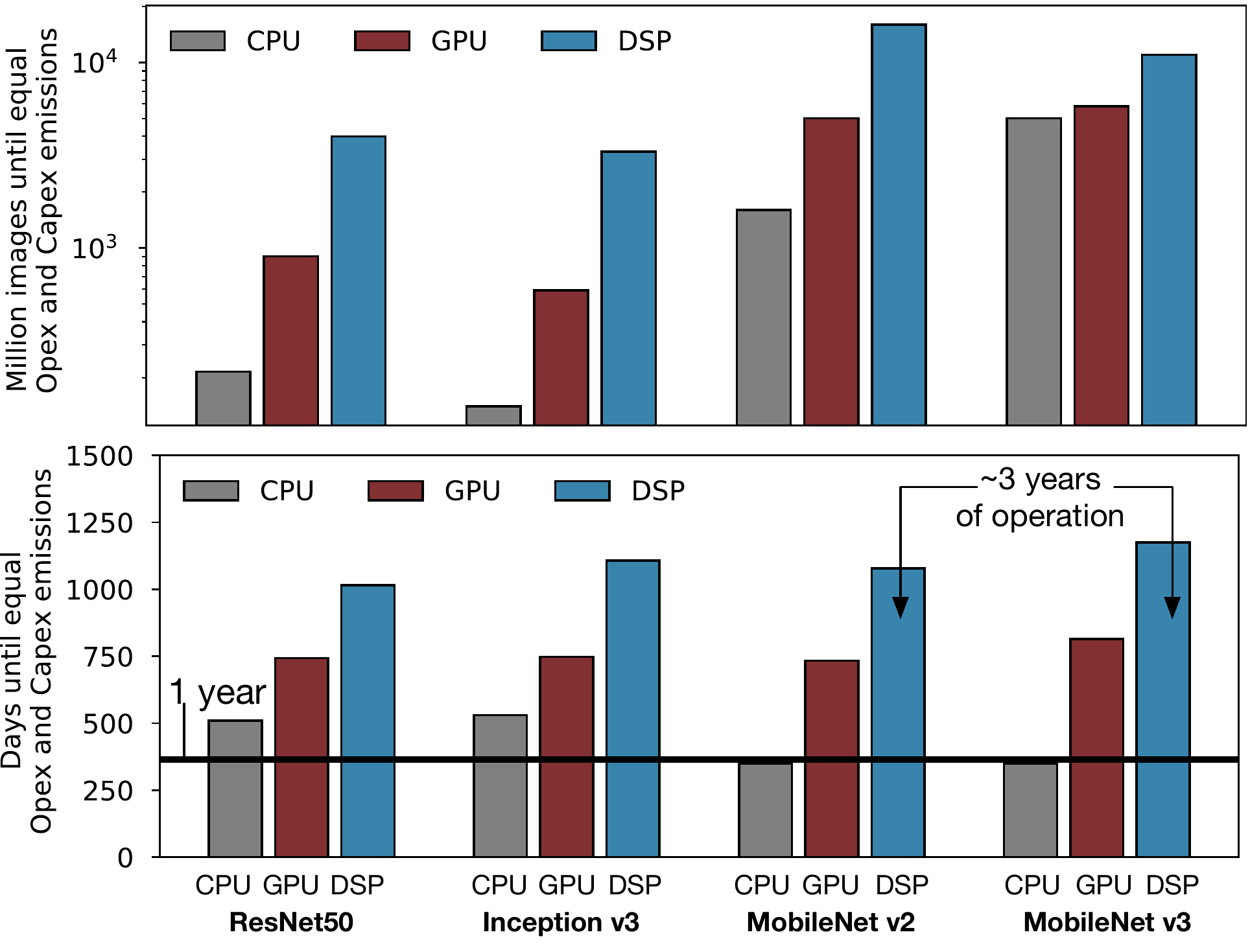}
  \vspace{-1em}
  \caption{Evaluating carbon footprint between manufacturing- and operational-related activities for Google Pixel 3 smartphone. 
  Algorithmic AI and hardware advances dramatically shifted carbon emissions toward manufacturing overhead.
  The top chart shows the number of inference images necessary for operational output to equal the integrated-circuit-manufacturing output. 
  The bottom chart shows how many days of image processing is necessary for operational output to equal integrated-circuit-manufacturing output.
  }
  \vspace{-1.0em}
  \label{fig:energy_pixel3}

\end{figure}

Carbon emissions from hardware manufacturing can be amortized by lengthening the hardware's operating time.
Here, we define the starting point of this amortization when the carbon output from operational use equals that from hardware manufacturing (i.e., the ratio of opex emissions to capex emissions is 1). 
Figure~\ref{fig:energy_pixel3} shows this breakeven in terms of the number of inferences (top) and days of operation (bottom) on a Google Pixel 3 phone.
We converted our measured power consumption (using a Monsoon power monitor~\cite{msoon}) to operational carbon emissions by assuming the average US energy-grid output: 380~g of \co~per kilowatt-hour~\cite{henderson2020climate}.
Finally, the manufacturing carbon footprint considers the overhead of building the SoC alone---assuming half the production carbon emissions are due to integrated circuits (see Figure~\ref{fig:apple_breakdown}). 

Algorithmic and architectural innovation has boosted energy efficiency, lengthening the amortization time. 
For instance, Figure~\ref{fig:energy_pixel3} (top) shows that to bring the carbon output from energy consumption (opex) to parity with that from hardware manufacturing (capex), ResNet-50 requires 200 million images and Inception v3 requires 150 million images when running on a CPU~\cite{he2016deep,szegedy2015going}. MobileNet v3 on a CPU takes five billion~\cite{howard2019searching} images; the 25$\times$ increase owes to algorithmic advances reducing mobile AI's memory and compute requirements.
In addition, hardware enhancements also reduce operational emissions.
For example, running MobileNet v3 on a DSP rather than a CPU reduces the operational footprint by 2$\times$, requiring 10 billion images for operational- and hardware-manufacturing-related carbon emissions to balance.
In comparison, the ImageNet training set consists of 14 million images~\cite{russakovsky2015imagenet}.

Furthermore, Figure~\ref{fig:energy_pixel3} illustrates how many days of continual AI inference are necessary for the operational carbon footprint to equal the hardware-manufacturing footprint.
MobileNet v3 running on a CPU, for example, takes 350 days of continuous operation.
DSPs increase the duration to nearly 1,200 days due to 1.5$\times$ and 2.2$\times$ improvements in performance and power efficiency, respectively.
By comparison, the device's expected lifetime is three years (about 1,100 days). 
Generally, given algorithmic and architectural enhancements, amortizing carbon emissions from hardware manufacturing requires performing AI inference beyond the expected lifetime of most mobile devices. 


\begin{figure}[t!]
  \centering
  \includegraphics[width=\columnwidth]{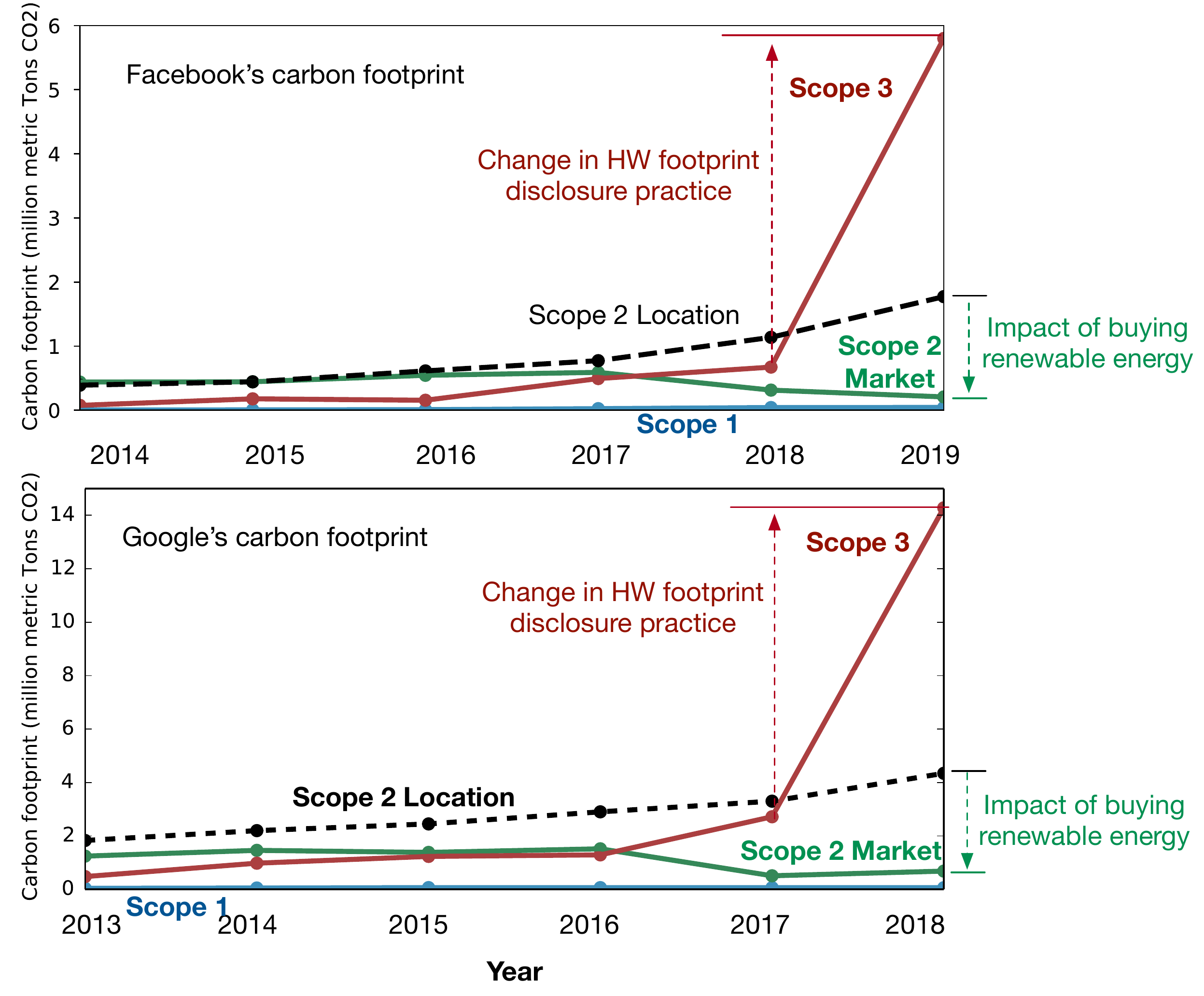}
  \vspace{-2em}
  \caption{ Carbon footprint of Facebook and Google (two large data center operators).
  As data centers increasingly rely on renewable energy, carbon emissions originate more from Scope 3, or supply-chain emissions (e.g., hardware manufacturing and construction).}
  \vspace{-1em}
  \label{fig:datacenter_timeline}

\end{figure}

\section{Environmental impact of data centers}~\label{sec:datacenters}
As AI, autonomous driving, robotics, scientific computing, AR/VR, and other emerging applications become ubiquitous, considering the environmental implications of both edge and data-center systems becomes important.
In this section we explore the environmental impact of data centers.
First we consider the carbon-emission breakdown of Facebook and Google facilities using industry-reported GHG Protocol data.
Next, we discuss the historical trends of data-center carbon emissions.
Our discussion highlights the positive impact of renewable energy on these emissions and the need for more-detailed accounting and reporting.
Finally, we present a case study of renewable energy's effect on data-center footprint.

\subsection{Breakdown of warehouse-scale data centers}
\textbf{Takeaway 7:} \textit{For data-center operators and cloud providers, most emissions are capex-related---for example, construction, infrastructure, and hardware manufacturing.}

Figure~\ref{fig:datacenter_timeline} illustrates the carbon footprint of Google (2013 to 2018) and Facebook (2014 to 2019)~\cite{googSustainabilityReport, fbSustainabilityReport}.
Following the GHG Protocol, we split emissions into Scope 1 (blue), Scope 2 (green), and Scope 3 (red).
Recall that Scope 1 (opex) emissions come from facility use of refrigerants, natural gas, and diesel; Scope 2 (opex) emissions come from purchased electricity; and Scope 3 (capex) emissions come from the supply chain, including employee travel, construction, and hardware manufacturing (see Section~\ref{sec:quantify_carbon} for details).

Analyzing the most recent data, Scope 3 comprises the majority of emissions for both Google and Facebook. 
In 2018, Google reported 21$\times$ higher Scope 3 emissions than Scope 2 emissions---that is, 14,000,000 metric tons of \co versus 684,000.
In 2019, Facebook reported 23$\times$ higher Scope 3 emissions than Scope 2 emissions---that is, 5,800,000 metric tons of \co versus 252,000.

\begin{figure}[t!]
  \centering
  \includegraphics[width=0.7\columnwidth]{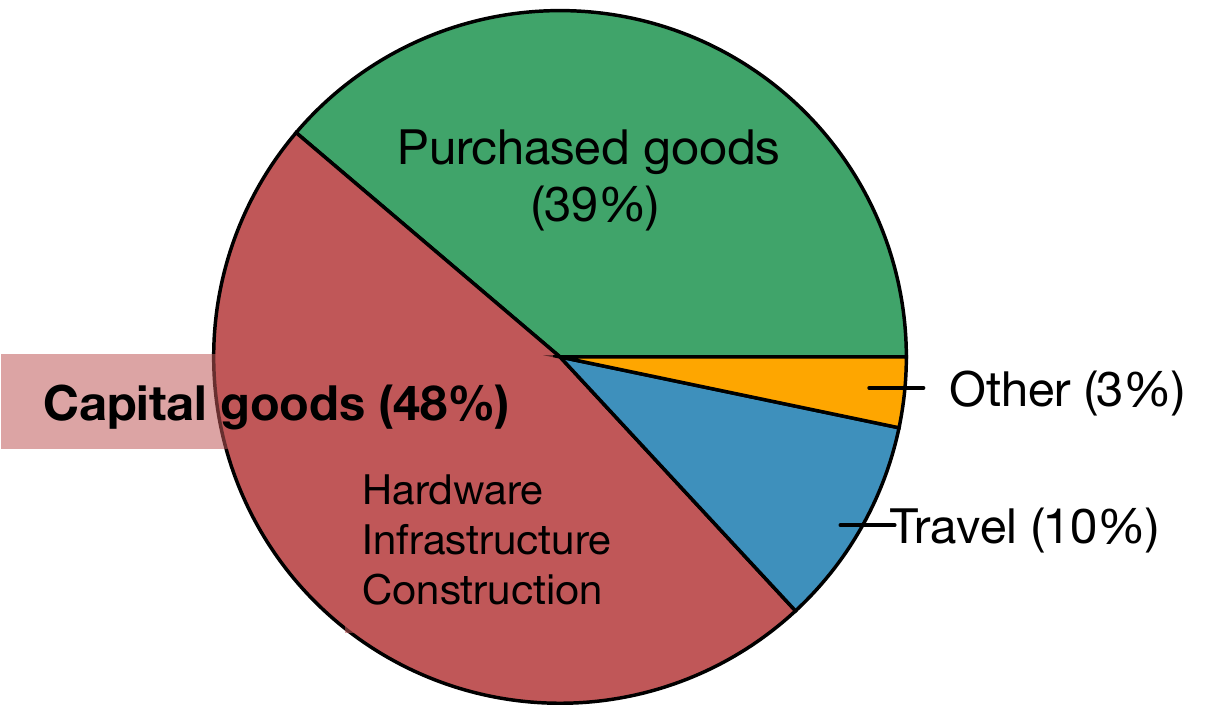}
  \vspace{-0.5em}
  \caption{ Breakdown of Facebook's 2019 Scope 3 carbon emissions.
  Capital goods (e.g., hardware, infrastructure, and construction) account for up to 48\% of the annual total.}
  \label{fig:fb_2019}

\end{figure}

Recall that Scope 3 emissions aggregate the entire supply chain; a large fraction of them are from data-center capex overhead such as construction and hardware manufacturing. 
Figure~\ref{fig:fb_2019} illustrates Facebook's breakdown of Scope 3 emissions in 2019.
Here, construction and hardware manufacturing (capital goods) account for up to 48\% of the company's Scope 3 emissions.

Similarly, we anticipate most of Google's Scope 3 emissions are from construction and hardware manufacturing.
Figure~\ref{fig:datacenter_timeline} shows that between 2017 and 2018, the company reported a 5$\times$ increase in that output.
We attribute the  large increase to the additional accounting and disclosure of hardware-manufacturing emissions---during that time, data-center energy consumption only increased by 30\%~\cite{googSustainabilityReport}.
Given the additional disclosure, the proportion of capex-related emissions increases compared with opex-related emissions.
Note that because industry disclosure practices are evolving, publicly reported Scope 3 carbon output should be interpreted as a lower bound. 
The varying guidelines highlight the importance of better carbon accounting and reporting.


\subsection{Impact of renewable energy}
To decrease operational carbon emissions, data centers are increasingly employing renewable energy.
Here we detail the impact of renewable energy on overall emissions and the breakdown between opex- and capex-related factors. 

\textbf{Takeaway 8:} \textit{
Although overall data-center energy consumption has risen over the past five years, carbon emissions from operational energy consumption have fallen.
The primary factor contributing to the growing gap between data-center energy consumption and carbon output is the use of renewable energy.} 

Figure~\ref{fig:datacenter_timeline} illustrates the carbon footprint of Google and Facebook over six years.
Although the figure divides these emissions into Scope 1, Scope 2, and Scope 3, Scope 2 comprises two types: location based and market based.
Location-based emissions assume the local electricity grid produces the energy---often through a mix of brown (i.e., coal and gas) and green sources.
Market-based emissions reflect energy that companies have purposefully chosen or contracted---typically solar, hydroelectric, wind, and other renewable sources.
Around 2013, Facebook and Google began procuring renewable energy to reduce operational carbon emissions.
These purchases decreased their operational carbon output even though their energy consumption continued to increase.
Thus, minimizing the emissions related to data-center workloads and hardware must consider renewable energy and the tradeoffs between opex- and capex-related factors.


\textbf{Takeaway 9:} \textit{For hardware, the carbon footprint between hardware manufacturing and use depends on the energy source.
Powering hardware with renewable energy reduces emissions from operational energy consumption; consequently, hardware manufacturing begins to dominate the carbon footprint.}

\begin{table}[t!]
\begin{center}
\small
\begin{tabular}{|c|c|c|c}
\hline
\multirow{2}{*}{\textbf{Source}} & Carbon intensity & Energy-payback \\ 
& (g \co/kWh) & time (months) \\ \hline \hline 
\textbf{Coal} & 820 & 2~\cite{weissbach2013energy} \\ \hline
\textbf{Gas} & 490 & 1~\cite{weissbach2013energy} \\  \hline
\textbf{Biomass} & 230 & $\sim$12~\cite{madsen2018carbon} \\ \hline
\textbf{Solar} & 41 & $\sim$36~\cite{nrealpvepbt}\\ \hline
\textbf{Geothermal} & 38 & 72~\cite{li2015comparison}\\ \hline
\textbf{Hydropower} & 24 & $\sim$12--36~\cite{weissbach2013energy,hyrdoIndia} \\ \hline 
\textbf{Nuclear} & 12 & 2~\cite{weissbach2013energy} \\ \hline 
\textbf{Wind} & 11 & $\leq$12~\cite{bonou2016life} \\ \hline \hline

\end{tabular}
\end{center}
\caption{Carbon efficiency of various renewable-energy sources.}
\vspace{-1em}
\label{tab:energy}
\end{table}

\begin{table}[t!]
\begin{center}
\small
\begin{tabular}{|c|c|c|}
\hline
\textbf{Geographic} & Carbon intensity & Dominant  \\ 
\textbf{average} & (g \co~/ kWh) & energy source \\ \hline \hline 
\textbf{World } & 301 & -- \\ \hline
\textbf{India } & 725 & Coal/gas\\ \hline
\textbf{Australia } & 597 & Coal \\ \hline
\textbf{Taiwan } & 583 & Coal/gas\\ \hline
\textbf{Singapore } & 495 & Gas \\ \hline
\textbf{United States } & 380 &  Coal/gas\\  \hline 
\textbf{Europe } & 295 & -- \\  \hline 
\textbf{Brazil } & 82 &  Wind/hydropower\\  \hline 
\textbf{Iceland } & 28 & Hydropower\\ \hline

\end{tabular}
\end{center}
\caption{Global carbon efficiency of energy production~\cite{electricityMap,henderson2020climate,indiapower}.}
\vspace{-3em}
\label{tab:energy_geo}
\end{table}

\begin{figure}[t!]
  \centering
  \includegraphics[width=0.93\columnwidth]{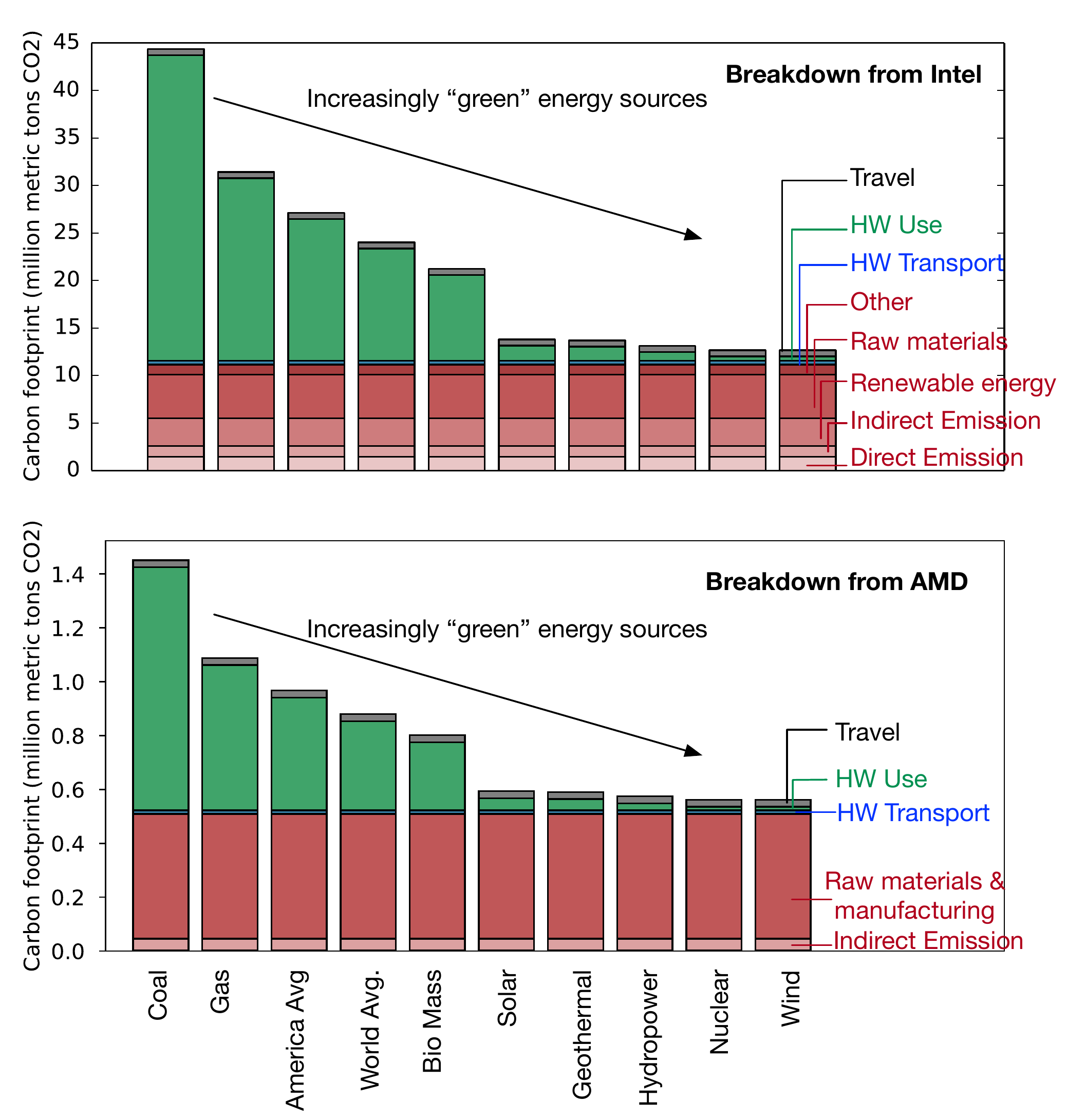}
  \vspace{-1em}
  \caption{ Reported carbon-footprint breakdown for Intel (top) and AMD (bottom) as renewable energy increasingly (from left to right) powers hardware operation.
  The use of renewable energy reduces carbon emissions dramatically; most of the remaining emissions are from manufacturing.}
  \vspace{-1.0em}
  \label{fig:intel_breakdown}

\end{figure}

Figure~\ref{fig:intel_breakdown} illustrates the impact of renewable energy on opex- and capex-related emissions on the basis of reported carbon data from Intel (top) and AMD (bottom)~\cite{intelSustainabilityReport, amdSustainabilityReport}.
The format mimics hardware life cycles.
Carbon emissions from device use over a three-year lifetime appear in green; emission from hardware manufacturing appear in red.
Although Intel and AMD both assume the average US energy-grid mix, we scaled the hardware-use emissions in accordance with variable energy sources.
Recall that warehouse-scale data centers often employ solar, hydropower, wind, and other renewable-energy sources.

The use of renewable energy dramatically changes the breakdown of carbon emissions across the hardware life cycle. 
For the baseline, assuming the US energy grid, roughly 60\% of Intel's carbon emissions and 45\% of AMD's come from hardware use and energy consumption.
With renewable energy, however, emissions from operational consumption decrease.
This decline is because renewable energy is orders of magnitude more efficient in grams of \co~emitted per kilowatt-hour of energy generated, as Table~\ref{tab:energy} shows.
Table~\ref{tab:energy_geo} lists the carbon intensity of energy production globally.
The baseline case assumes the US energy grid (301 g \co~per kilowatt-hour); solar and wind emit 41 g and 11 g of \co~per kilowatt-hour, respectively.
Figure~\ref{fig:intel_breakdown} shows that when using solar and wind, which frequently power data centers, over 80\% of emissions come from hardware manufacturing.

Designing sustainable data centers should therefore consider the role of renewable energy, the effect of efficiency increases on opex-related emissions, and the effect of resource provisioning and leaner hardware on capex-related emissions.




\section{Environmental impact from manufacturing}~\label{sec:manufacturing}
So far, our results show hardware manufacturing comprises a large portion of emissions in both mobile and data-center systems.
In data centers, renewable energy is a significant contributor to the opex-related footprint.
In this section, we consider the carbon footprint of chip manufacturing and the impact of powering fabs using renewable energy.

\textbf{Takeaway 10}: \textit{Using renewable energy to power fabs will reduce the carbon emissions from hardware manufacturing. 
Even under optimistic renewable-energy projections, however, manufacturing will continue to represent a large portion of hardware-life-cycle carbon footprints.}

\begin{figure}[t!]
  \centering
  \includegraphics[width=0.95\columnwidth]{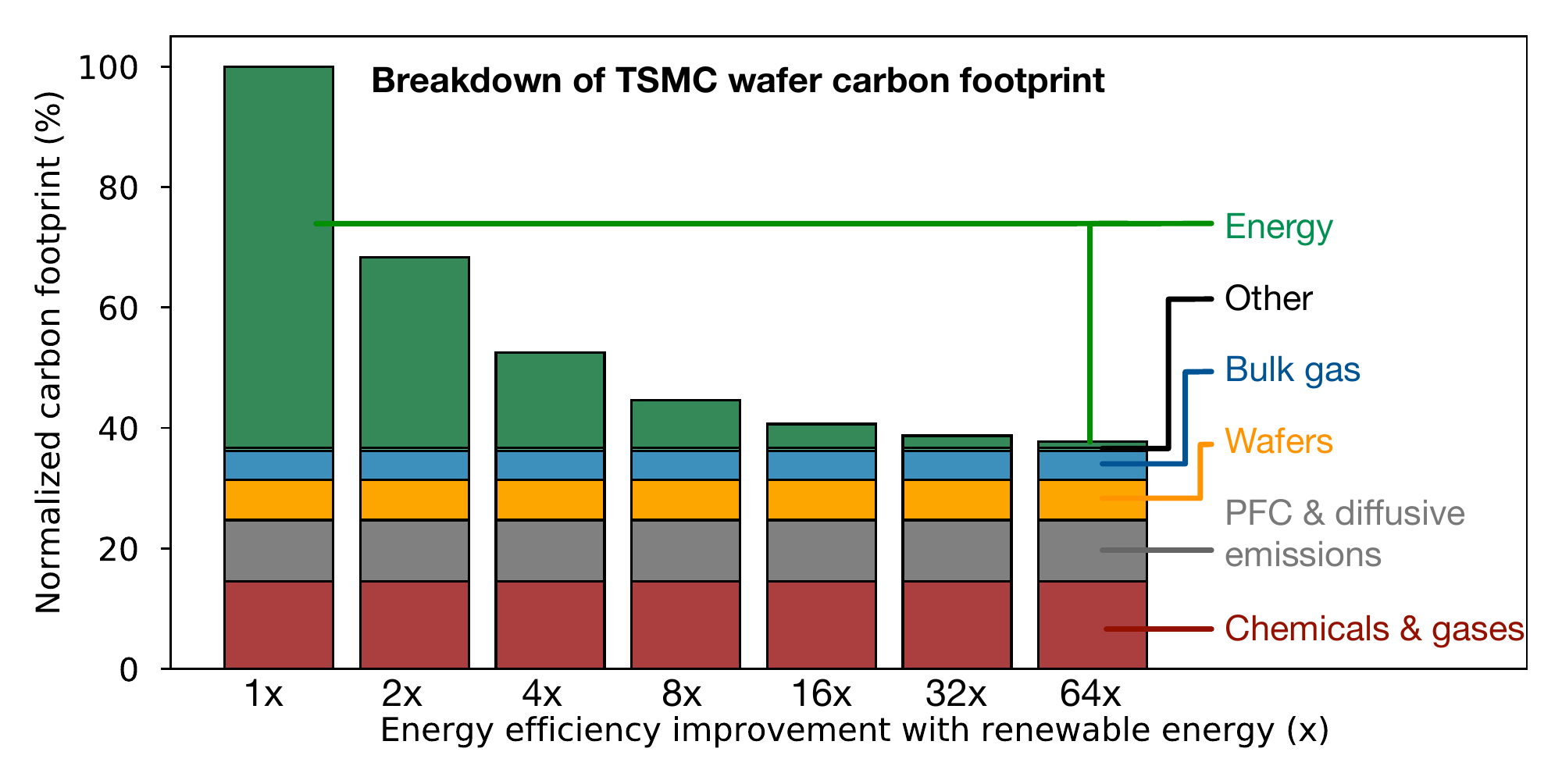}
  \vspace{-1em}
  \caption{ Carbon-emissions breakdown for TSMC wafer manufacturing.
            Renewable energy provides up to a 64$\times$ reduction in emissions from electricity, and overall emissions for wafers drops by 2.7$\times$.
            Although the reduction will reduce the carbon output of manufacturing, consideration of capex-related emissions for mobile and data-center hardware will remain important.}
  \label{fig:tsmc}
  \vspace{-1em}
\end{figure}

Figure~\ref{fig:tsmc} shows the carbon breakdown for wafer manufacturing at TSMC~\cite{tsmcSustainabilityReport}.
The breakdown is normalized to the baseline energy source.
To model the impact of renewable energy, we vary the carbon intensity of the energy consumed.
Although the precise energy-grid efficiency is unknown, our analysis considers a range of improvements, including the best case: replacing coal with 100\% wind energy, for a 70$\times$ improvement (see Table~\ref{tab:energy}).
Using greener energy directly reduces the fab's carbon output from consumed energy (green).

Even though using renewable energy can cut a fab's hardware-manufacturing carbon emissions, minimizing life-cycle and hardware-manufacturing emissions will remain important.
As Figure~\ref{fig:tsmc} shows, a 64$\times$ boost in renewable energy reduces the overall carbon output by roughly 2.7$\times$, an ambitious goal.
By 2025, TSMC estimates renewable energy will produce 20\% of the electricity that drives forthcoming 3nm fabs~\cite{tsmcSustainabilityReport}.
Intel already uses renewable energy to meet much of its demand; only 9.7\% of the energy consumed by Intel fabs comes from nonrenewable sources (see Figure~\ref{fig:intel_breakdown}).
Recall that roughly 75\% of the carbon footprint for battery-powered devices is from hardware manufacturing (see Section~\ref{sec:mobile}), and opex is a small fraction for data centers.
Even as fabs employ more renewable energy to reduce their environmental impact, hardware manufacturing will remain an important aspect of designing sustainable computers and workloads. 

\section{Addressing carbon footprint of systems}~\label{sec:future_direction}
\begin{figure}[t!]
  \centering
  \includegraphics[width=\columnwidth]{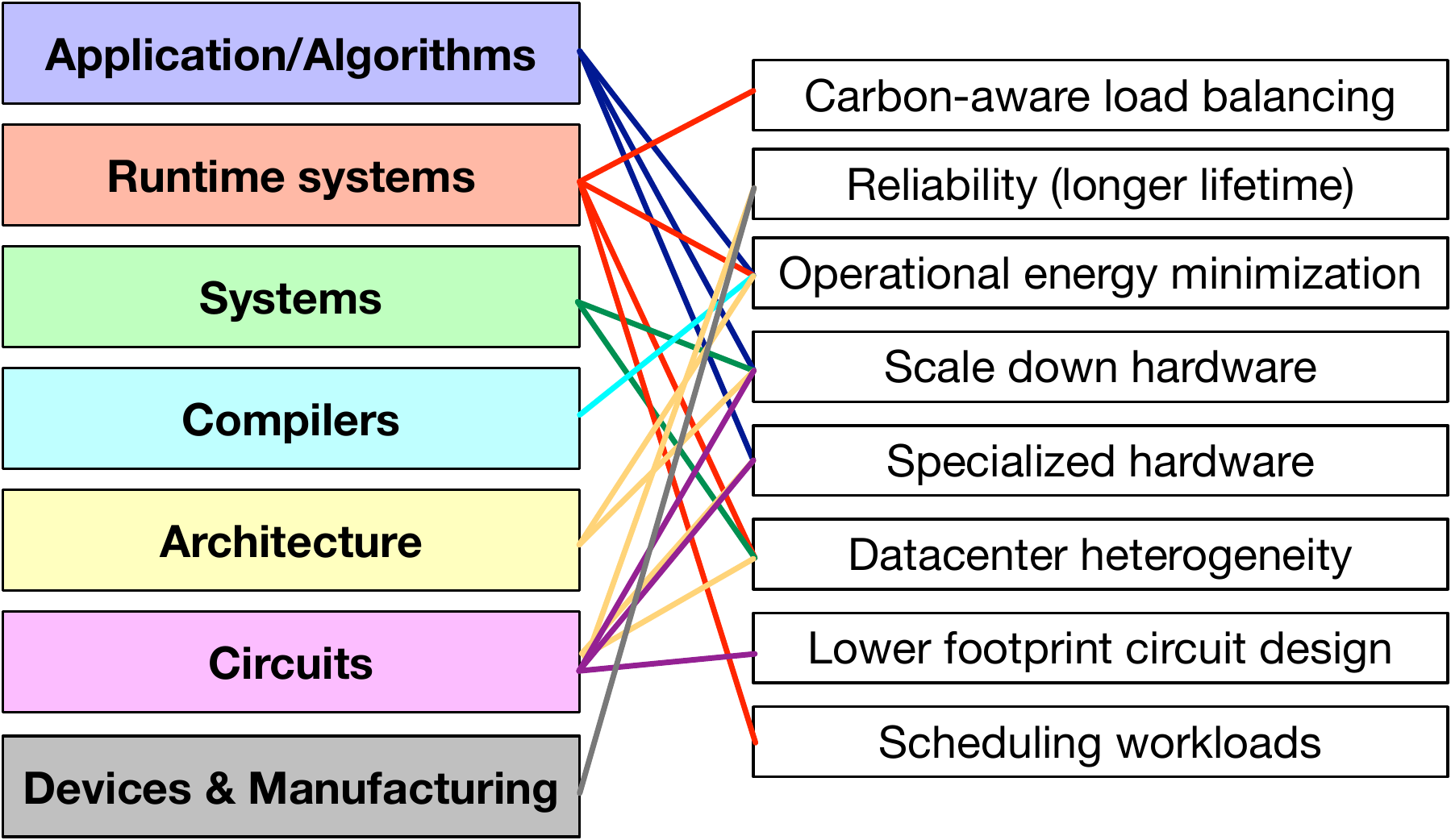}
  \caption{ Reducing the carbon output of computer systems requires cross-layer optimization across the computing stack. The potential opportunities (right) overlap with multiple stack layers (left). }
  \vspace{-1.0em}
  \label{fig:computing_stack}

\end{figure}

Optimizing the environmental impact of mobile and data-center computing platforms requires addressing the carbon footprint from operational energy consumption (opex) and hardware manufacturing (capex).
Given its immediate importance and scale, we must adopt vertically integrated research methods to minimize the emissions associated with computing. 
Figure~\ref{fig:computing_stack} illustrates some of the important research directions and their corresponding layers in the computing stack. 
This section outlines future directions across the computing stack in light of the state-of-the-art and prior work.
We use AI training and inference as an example application, but the tradeoffs extend to other domains. 












\textbf{Applications and algorithms.}
Application- and algorithm-level optimizations can reduce both opex- and capex-related carbon emissions.
As a motivating example, consider AI training.
The parameters of the energy footprint for training AI models are threefold: the footprint of processing one example ($E$), the data-set size ($D$), and the hyperparameter search ($H$)~\cite{schwartz2019green, strubell2019energy}.
Even though some data centers employ renewable energy (see Section~\ref{sec:datacenters}), researchers must still consider the hardware's carbon footprint.

Table~\ref{tab:co_amd} presents the compute, memory, and carbon footprint characteristics for two Apple Mac Pro desktops.
Compared with the first configuration, the second has dual AMD Radeon Vega GPUs, providing 4$\times$, 8$\times$, and 16$\times$ more flops, memory bandwidth, and capacity, respectively. 
It represents a data-center-scale server~\cite{gupta2020architectural, gupta2020deeprecsys}, yielding a 2.6$\times$ greater manufacturing carbon footprint. 

Reducing carbon emissions requires training on systems with fewer resources.
Given the faster--than--Moore's Law AI improvement, novel methods to train models given lesser compute and storage capabilities can directly reduce associated carbon emissions (i.e., $E$, $D$).
Similarly, reducing the hyperparameter-search factor ($H$) reduces the necessary number of parallel training nodes and the AI-training carbon footprint~\cite{hernandez2020measuring, schwartz2019green}.
Generally, algorithmic optimizations for scale-down systems will drastically cut emissions. 



\begin{table}[t!]
\begin{center}
\small
\begin{tabular}{|c|c|c|}
\hline
\textbf{Parameter} & \textbf{Mac Pro 1}  & \textbf{Mac Pro 2}  \\ \hline \hline
CPU (cores $\times$ threads) & 8$\times$2 & 28$\times$2 \\ \hline
DRAM (GB) & 32 & 1,536 \\ \hline
Storage (GB) & 256 & 4,096 \\ \hline

GPU performance (teraflops) & 6.2 & 28.4 \\ \hline
GPU-memory BW  (GB/s) & 256 & 2,048 \\ \hline
System TDP (W) & 310 & 730 \\ \hline \hline

Manufacturing \co~(kg) & 700 & 1,900 \\ \hline 
\end{tabular}
\end{center}
\caption{Comparing Apple Mac Pro desktops.
The high-performance configuration---more cores, memory, storage, GPU flops, and GPU memory bandwidth---has a 2.7$\times$ higher manufacturing-related \co~\cite{MacPro}.}
\label{tab:co_amd}
\vspace{-3em}
\end{table}

\textbf{Run-time systems.} 
Run-time systems, including schedulers, load-balancing services, and operating systems, can reduce both opex- and capex-related carbon footprints. 
Optimizing for opex-related emissions, cloud providers use machine learning to improve the efficiency of data-center cooling infrastructure~\cite{google_2020}.
Furthermore, recent work proposes scheduling batch-processing workloads during periods when renewable energy is readily available~\cite{carbonawareloadbalancing, le2010capping, ZhenhuaRenewable, li2011characterizing, gujarati2017swayam}. 
Doing so decreases the average carbon intensity (see Table~\ref{tab:energy}) of energy consumed by data-center services.

Optimizing for capex-related emissions requires reducing hardware resources.
Recently proposed schedulers optimize infrastructure efficiency in terms of total power consumption while balancing performance for latency-critical and batch-processing workloads~\cite{kanev2014tradeoffs,wu2016dynamo,sriraman2019softsku, gupta2020architectural, gupta2020deeprecsys}.
Others have proposed novel scheduler designs to enable energy-efficient, collaborative cloud and edge execution~\cite{kim2020autoscale, kang2017neurosurgeon}.
Our analysis enables future studies to co-optimize for tail-latency-, throughput-, power-, and infrastructure-related carbon emissions.

\textbf{Systems.}
Systems researchers can guide overall mobile- and data-center-scale system provisioning to directly reduce capex-related emissions.
Recently, systems have scaled up and out to boost performance and power efficiency~\cite{magaki2016asic,grot2012optimizing}. 
Figure~\ref{fig:phone_perf_cf_pareto} and Figure~\ref{fig:energy_pixel3a} show more-sophisticated hardware for improving mobile-AI performance and power efficiency; simultaneously, scale-up hardware increases manufacturing- and device-level carbon emissions. 
Instead, as Table~\ref{tab:co_amd} shows, scaling systems down can reduce the environmental impact of hardware---assuming they still provide sufficient performance.

Data centers often comprise heterogeneous platforms, with custom hardware for important applications, to maximize infrastructure efficiency (e.g., performance and power). 
For instance, Facebook data centers have custom servers for AI inference and training~\cite{hazelwood2018applied}.
Our work enables systems researchers to consider how heterogeneity can reduce carbon footprint by reducing overall hardware resources in the data center. 
But researchers must balance the environmental benefits with the challenges of heterogeneity: programmability, resource management, cross-ISA execution, and debugging~\cite{sigarchBlogDelimitrou}.

\textbf{Compilers and programming languages.}
Recent work has proposed new programming languages to enable more-energy-efficient code~\cite{sampson:pldi2011}. 
Others propose compiler-level optimizations to increase the code's energy efficiency~\cite{schulte:asplos14,Wu:micro2005,Wu:ieeemicro2005}. 
Although optimizing nonfunctional properties, such as minimizing energy consumption through instruction-level energy annotation~\cite{Shao:islped2013,Pandiyan:iiswc2014,Arunkumar:hpca2019}, can indirectly mitigate environmental impact, future compiler and programming-language technologies can optimize for \co~emissions directly by applying our emission analysis.   

\textbf{Architecture.} 
Over the past 20 years, computer-architecture researchers and designers have devoted substantial effort to energy-efficient mobile and data-center systems.
Their work includes system-performance increases, power management and energy-efficiency optimization through dynamic voltage and frequency scaling (DVFS) ~\cite{Shingari:ispass2018,Wang:dac2011,Qiu:dac99,Gaudette:tmc2019,Gaudette:hpca2016,Wu:micro2005,Wu:ieeemicro2005,Isci:micro2006a,Isci:micro2006b,Buyuktosunoglu:pacs2000,Srinivasan:micro2002,Rangan:isca2009,Teodorescu:isca2008}, and specialized hardware~\cite{eie,eyeriss,tpu,minerva,cambricon-x,zhang2016cambricon}. 
As Figure~\ref{fig:energy_pixel3a} and Figure~\ref{fig:energy_pixel3} show, architectural improvements have considerably reduced the operational carbon footprint of mobile AI inference.

Future architecture research can also minimize capex-related carbon emissions.
For instance, as Table~\ref{tab:co_amd} shows, higher-performance hardware incurs higher manufacturing-related carbon emissions.
More generally, as billion-transistor devices experience low utilization, systems must balance dark silicon with manufacturing emissions~\cite{esmaeilzadeh2012dark}.
Similarly, architectural optimizations can directly reduce \co~output by judiciously provisioning resources, scaling down hardware, and incorporating specialized circuits. 

\textbf{Circuits.}
In addition to architectural innovations, circuit designers have enabled high-performance and low-power hardware through efforts that include clock/power gating~\cite{hu2004microarchitectural}, DVFS~\cite{kim2008system}, and circuit-level timing speculation~\cite{ernst2003razor}.
These efforts indirectly minimize opex-related carbon emissions.

Future circuit research can also reduce capex-related carbon emissions.
First, it may consider circuit-level resource provisioning to balance performance, area, energy efficiency, and carbon footprint.
Next, DRAM and NAND-flash-memory research should investigate low-carbon technologies and higher reliability to lengthen hardware lifetimes.
Finally, vertically integrated research into specializing low-carbon circuits for salient applications will also decrease capex-related emissions.
For example, in the case of AI, co-designing neural-network fault tolerance and compression for specialized circuits and memories can allow hardware consolidation and, therefore, smaller carbon footprints~\cite{pentecost2019maxnvm, minerva, eyeriss, gupta2019masr, eie}. 

%

\textbf{Semiconductor devices and manufacturing.} Finally, capex emissions must be addressed through device modeling, characterization, design, and fab manufacturing. 
For instance, hardening a device's reliability and endurance extends its lifetime, cutting capex-related carbon emissions. 
Moreover, research into sustainable manufacturing processes via novel devices, yield enhancement, fabrication materials, renewable-energy sources, and maximum operating efficiency will directly reduce production overhead. 

\section{Conclusion and future work}
As computing technology becomes ubiquitous, so does its environmental impact.
This work shows how developers and researchers should approach the environmental consequences of computing, from mobile to data-center-scale systems.
First, we demonstrated that reducing energy consumption alone fails to reduce carbon emissions.
Next, we described the industry's practice for quantifying the carbon output of organizations and individual systems. 
Finally, on the basis of our analysis, we characterized the carbon emissions of various hardware platforms.
Our effort demonstrates that over the last decade, hardware manufacturing---as opposed to operational energy consumption---has increasingly dominated the carbon footprint of mobile systems.
Similarly, as more data centers employ renewable energy, the dominant source of their total carbon footprint becomes hardware manufacturing.  

We hope this work lays the foundation for future investigation of environmentally sustainable systems.
Designing, building, and deploying such systems requires collective industry/academic collaboration.
We conclude by outlining future steps toward that goal.

\textbf{Better accounting practices.} Although many organizations publicly report their carbon emissions, improved accounting (e.g., broader participation as well as standardized accounting and disclosures) will provide further guidance on tackling salient challenges in realizing environmentally sustainable systems.

\textbf{Carbon footprint as a first-order optimization target.} 
Researchers and developers across the computing stack should consider carbon footprint to be a first-class design metric alongside increased workload and system characterization, and they should incorporate optimizations for lower environmental impact.




\textbf{Beyond carbon emissions.}
Although this work focuses on carbon emissions, the environmental impact of computing systems is multifaceted, spanning water consumption as well as use of other natural resources, including aluminum, cobalt, copper, glass, gold, tin, lithium, zinc, and plastic.


\section{Acknowledgements}

We would like to thank Urvi Parekh, Stephanie Savage, and Julia Rogers for their valuable insights and many discussions on the environmental sustainability of technology companies.
The collaboration leads to the deeper understanding of the challenges technology companies face in enabling environmentally sustainable operation presented in this work.
We would also like to thank Kim Hazelwood for supporting and encouraging this work.
The support helped foster new interdisciplinary collaborations on understanding and tackling the environmental impact of computing.


\bibliographystyle{ieeetr}
\bibliography{references}

\end{document}